\documentclass[aps,prl,reprint,superscriptaddress]{revtex4-1}
\usepackage{graphicx}
\usepackage{color,amsmath}

\usepackage[normalem]{ulem}

\begin{document}


\title{Giant spin-orbit splitting in inverted InAs/GaSb double quantum wells}



\author{Fabrizio Nichele}
\email[email: ]{fnichele@nbi.ku.dk}
\affiliation{Center for Quantum Devices and Station Q Copenhagen, Niels Bohr Institute, University of Copenhagen, Universitetsparken 5, 2100 Copenhagen, Denmark}

\author{Morten Kjaergaard}
\affiliation{Center for Quantum Devices and Station Q Copenhagen, Niels Bohr Institute, University of Copenhagen, Universitetsparken 5, 2100 Copenhagen, Denmark}

\author{Henri J. Suominen}
\affiliation{Center for Quantum Devices and Station Q Copenhagen, Niels Bohr Institute, University of Copenhagen, Universitetsparken 5, 2100 Copenhagen, Denmark}

\author{Rafal Skolasinski}
\affiliation{QuTech, Delft University of Technology, 2600 GA Delft, The Netherlands}

\author{Michael Wimmer}
\affiliation{QuTech, Delft University of Technology, 2600 GA Delft, The Netherlands}

\author{Binh-Minh Nguyen}
\affiliation{HRL Laboratories, 3011 Malibu Canyon Road, Malibu, California 90265, USA}

\author{Andrey A. Kiselev}
\affiliation{HRL Laboratories, 3011 Malibu Canyon Road, Malibu, California 90265, USA}

\author{Wei Yi}
\affiliation{HRL Laboratories, 3011 Malibu Canyon Road, Malibu, California 90265, USA}

\author{Marko Sokolich}
\affiliation{HRL Laboratories, 3011 Malibu Canyon Road, Malibu, California 90265, USA}

\author{Michael J. Manfra}
\affiliation{Department of Physics and Astronomy and Station Q Purdue, Purdue University, West Lafayette, Indiana 47907 USA}
\affiliation{School of Materials Engineering, Purdue University, West Lafayette, Indiana 47907 USA}
\affiliation{School of Electrical and Computer Engineering, Purdue University, West Lafayette, Indiana 47907 USA}
\affiliation{Birck Nanotechnology Center, Purdue University, West Lafayette, Indiana 47907 USA}

\author{Fanming Qu}
\affiliation{QuTech, Delft University of Technology, 2600 GA Delft, The Netherlands}

\author{Arjan J.A. Beukman}
\affiliation{QuTech, Delft University of Technology, 2600 GA Delft, The Netherlands}

\author{Leo P. Kouwenhoven}
\affiliation{QuTech, Delft University of Technology, 2600 GA Delft, The Netherlands}

\author{Charles M. Marcus}
\affiliation{Center for Quantum Devices and Station Q Copenhagen, Niels Bohr Institute, University of Copenhagen, Universitetsparken 5, 2100 Copenhagen, Denmark}


\date{\today}

\begin{abstract}
Transport measurements in inverted InAs/GaSb quantum wells reveal a giant spin-orbit splitting of the energy bands close to the hybridization gap. The splitting results from the interplay of electron-hole mixing and spin-orbit coupling, and can exceed the hybridization gap. We experimentally investigate the band splitting as a function of top gate voltage for both electron-like and hole-like states. Unlike conventional, noninverted two-dimensional electron gases, the Fermi energy in InAs/GaSb can cross a single spin-resolved band, resulting in full spin-orbit polarization. In the fully polarized regime we observe exotic transport phenomena such as quantum Hall plateaus evolving in $e^2/h$ steps and a non-trivial Berry phase.
\end{abstract}


\maketitle


The InAs/GaSb double quantum well (QW) shows a peculiar band alignment, with the InAs conduction band and the GaSb valence band residing very close in energy. Shifting the bands by tuning the QW thickness or applying perpendicular electric field yields a rich electronic phase diagram ~\cite{Yang1997,Liu2008,Qu2015,Nichele2015}.
When the InAs conduction band resides higher than the GaSb valence band, the band structure of a trivial insulator is obtained. By lowering the InAs conduction band below the GaSb valence band, a small hybridization gap opens at finite $k$-vectors \cite{Yang1997}.
Beyond topological-insulator behavior, expected to emerge in the hybridization gap \cite{Liu2008, Knez2011, Suzuki2013, Knez2014, Du2015, Mueller2015}, the impact of the inverted band structure on transport remains largely unexplored.

Here, we investigate experimentally and numerically how the combination of spin-orbit coupling (SOC) and electron-hole mixing results in a giant band splitting in InAs/GaSb heterostructures close to the hybridization gap. The two resulting subbands, with opposite spin-orbit eigenvalue and different carrier densities, contribute to transport in parallel, and can be detected via magnetotransport measurements. These results are of potential value to semiconductor spintronics, where two-dimensional electron gases (2DEGs) with sizable spin-orbit splittings at low density are desirable \cite{Zutic2004}. 

To quantify SOC directly from experimental data, without relying on any particular model, we use the spin-orbit polarization $(n_1-n_2)/(n_1+n_2)$, with $n_{1,2}$ the carrier densities of the split spin-orbit subbands \cite{Note_winkler}. In Rashba systems, the larger the SOC parameter $\alpha$, the larger the density difference of the subbands at the Fermi energy, with $\alpha$ typically increasing with density \cite{Winkler2003}. However, the spin-orbit polarization is usually smaller than $15\%$, even for 2DEGs with large SOC such as InAs, InSb or HgTe \cite{Luo1988,Das1989,Heida1998,Brosig1999,Gui2004,Nishioka2009}, while values up to $40\%$ are reached in GaAs or HgTe hole gases \cite{Stormer1983,Habib2004,Nichele2014b,Minkov2014}. In contrast, we find that the hybridized band structure of InAs/GaSb results in two striking peculiarities. First, the spin-orbit polarization increases approaching the charge neutrality point (CNP); second, the spin-orbit polarization reaches $100\%$.

Experiments were performed on a $12.5~\mathrm{nm}$ InAs, $5~\mathrm{nm}$ GaSb structure patterned in a $100\times 50 ~\mathrm{\mu m^2}$ Hall bar geometry oriented along the [110] crystallographic direction and covered with a global top gate. Magnetotransport measurements used conventional low-frequency lock-in techniques at a temperature of $50~\mathrm{mK}$. Additional information on the wafer structure, sample fabrication and measurement techniques are provided in the Supplemental Material~\cite{Supplement}.

\begin{figure}
\includegraphics[width=\columnwidth]{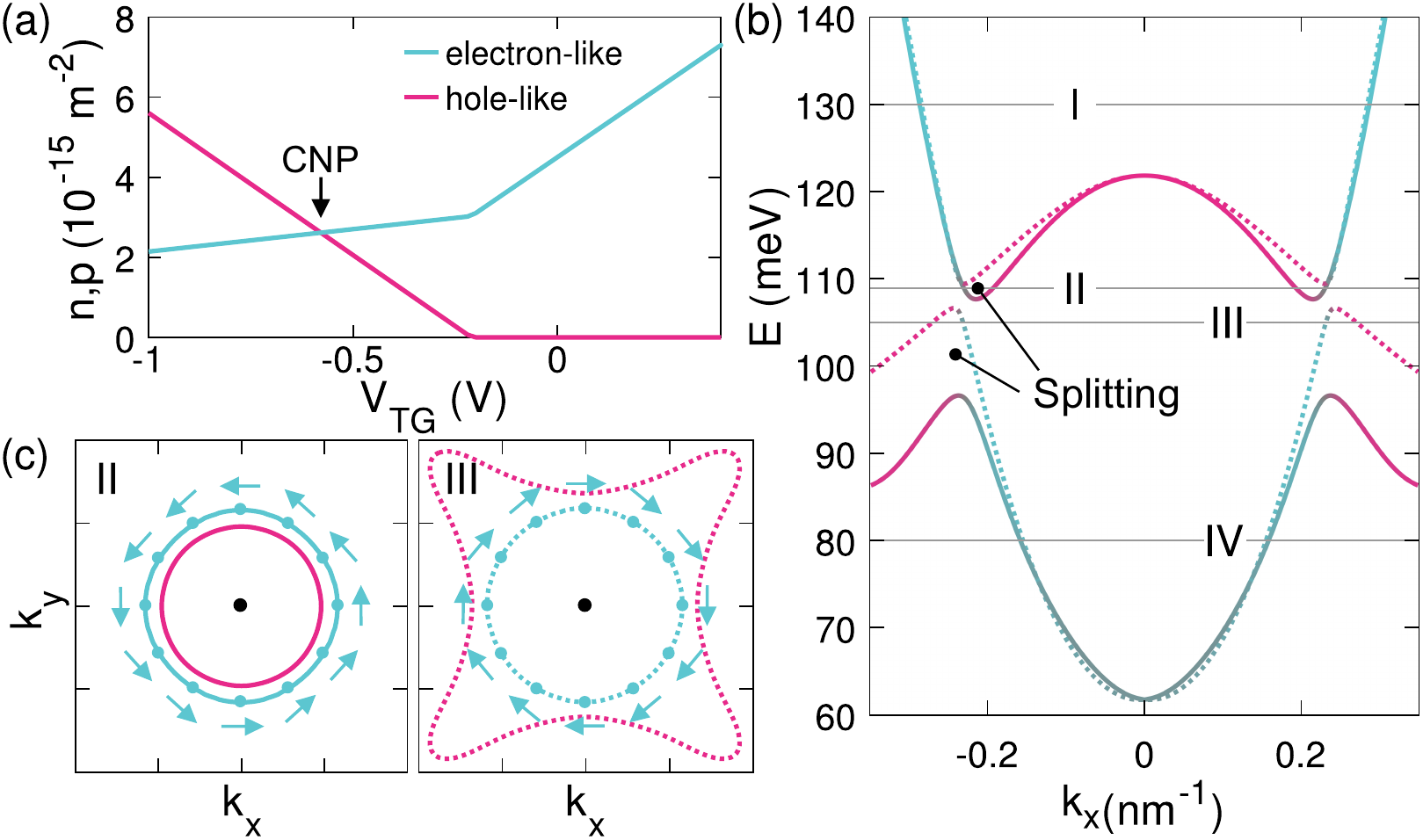}
\caption{(a) Expected electron and hole densities dependence on $V_{\mathrm{TG}}$. (b) Numerical band structure calculation for $V_{\mathrm{TG}}=-0.4~\mathrm{V}$. The color indicates the wavefunction main character, solid and dotted lines distinguish two spin-orbit split subbands. (c) Fermi contours and spin-texture of electron-like states for the Fermi energies II and III indicated in (b). The axis divisions are $0.2~\mathrm{nm^{-1}}$, with the black dot indicating the origin. Fermi pockets at large $k$-vector are ignored, but further discussed in the Supplemental Material \cite{Supplement}.}
\label{fig1}
\end{figure}

To realistically model our device, we first determine the band alignment as a function of top gate voltage, $V_{\mathrm{TG}}$, using a parallel plate capacitor model \cite{Qu2015} discussed in the Supplemental Material \cite{Supplement}. The model predicts the density dependence for electrons ($n$) and holes ($p$) shown in Fig.~\ref{fig1}(a). For $V_{\mathrm{TG}}>-0.2~\mathrm{V}$ only electrons are present in the system, with the kink in $n$ at $V_{\mathrm{TG}}=-0.2~\mathrm{V}$ coinciding with the onset of hole accumulation. Once the hole layer is populated, it partially screens the electrons from being further depleted via the top gate. The hybridization gap is expected at CNP, when $n=p$. The calculated electrostatic potential is then used for a $V_{\mathrm{TG}}$ dependent band structure simulation using standard $\boldsymbol{k}\cdot\boldsymbol{p}$ theory \cite{Supplement}. In particular, we are interested in the band structure of our system close to CNP.

The band structure for $V_{\mathrm{TG}}=-0.4~\mathrm{V}$ is presented in Fig.~\ref{fig1}(b). The band coloring represents the calculated wavefunction character (blue for electron-like and pink for hole-like states, also recognizable from the band curvature) while solid and dotted lines distinguish the spin-orbit species. In this configuration electron and hole bands are inverted and hybridized, with a small gap at finite $k$-vectors. Results for different gate voltages, shown in the Supplementary Material \cite{Supplement}, are qualitatively similar but with a varying band overlap.
Remarkably, SOC vertically splits the hybridized bands by a sizable amount resulting in a spin dependent hybridization gap. In this unique band structure, the Fermi energy can cross a \emph{single} branch of the spin split bands, as indicated by the energy levels II and III in Fig.~\ref{fig1}(b). In these situations the system contains both electron- and hole-like carriers, and the carriers of the same kind are \emph{fully spin-orbit polarized}. This effect is prominent close to the band crossing and negligible far from the hybridization gap [see I and IV in Fig.~\ref{fig1}(b)], as expected for individual InAs and GaSb QWs.
While the gap size and the bands overlap depend on $V_{\mathrm{TG}}$, the giant splitting at the CNP is a generic feature of the model. Qualitatively similar results were also obtained in previous calculations \cite{Zakharova2002, Liu2008, Li2009,Hu2016}. The simulation is consistent with our experiments, where we measure no clear gapped region at the CNP, but a giant spin-orbit splitting of electron- and hole-like states.

Fermi contours for energy levels II and III are shown in Fig.~\ref{fig1}(c), together with the calculated spin texture of electron-like states. The model indicates Rashba-like spin orientation with spins nearly perpendicular to the momentum direction, with small deviations due to the absence of axial symmetry. This situation is reminiscent of Dirac materials such as graphene or three-dimensional topological insulators, and signatures of Berry phase effects can be expected. Hole-like states are instead highly anisotropic.

\begin{figure}
\includegraphics[width=\columnwidth]{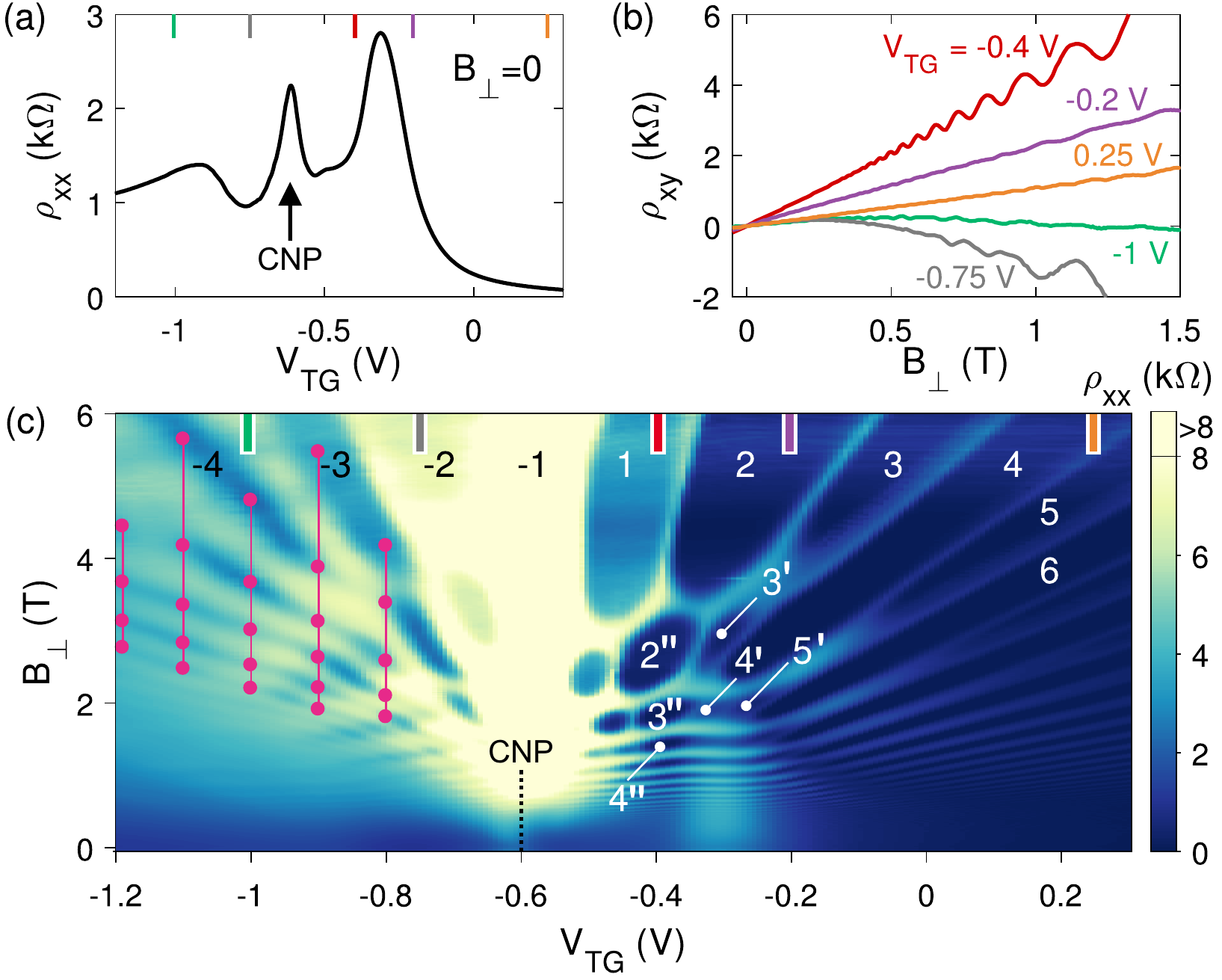}
\caption{(a) Longitudinal resistivity $\rho_{xx}$ as a function of top gate voltage for $B_\perp=0$, with the position of the charge neutrality point indicated. (b) Transverse resistivity $\rho_{xy}$ as a function of $B_\perp$ for different values of $V_{\mathrm{TG}}$, as also indicated by the markers in (a) and (c). (c) $\rho_{xx}$ as a function of $V_{TG}$ and $B_{\perp}$, with positive (negative) numbering indicating electron-like (hole-like) LLs. Pink dots denotes h-like filling factors and are used to extract the hole density shown in Fig.~\ref{fig3}(c).}
\label{fig2}
\end{figure}

\begin{figure}
\includegraphics[width=\columnwidth]{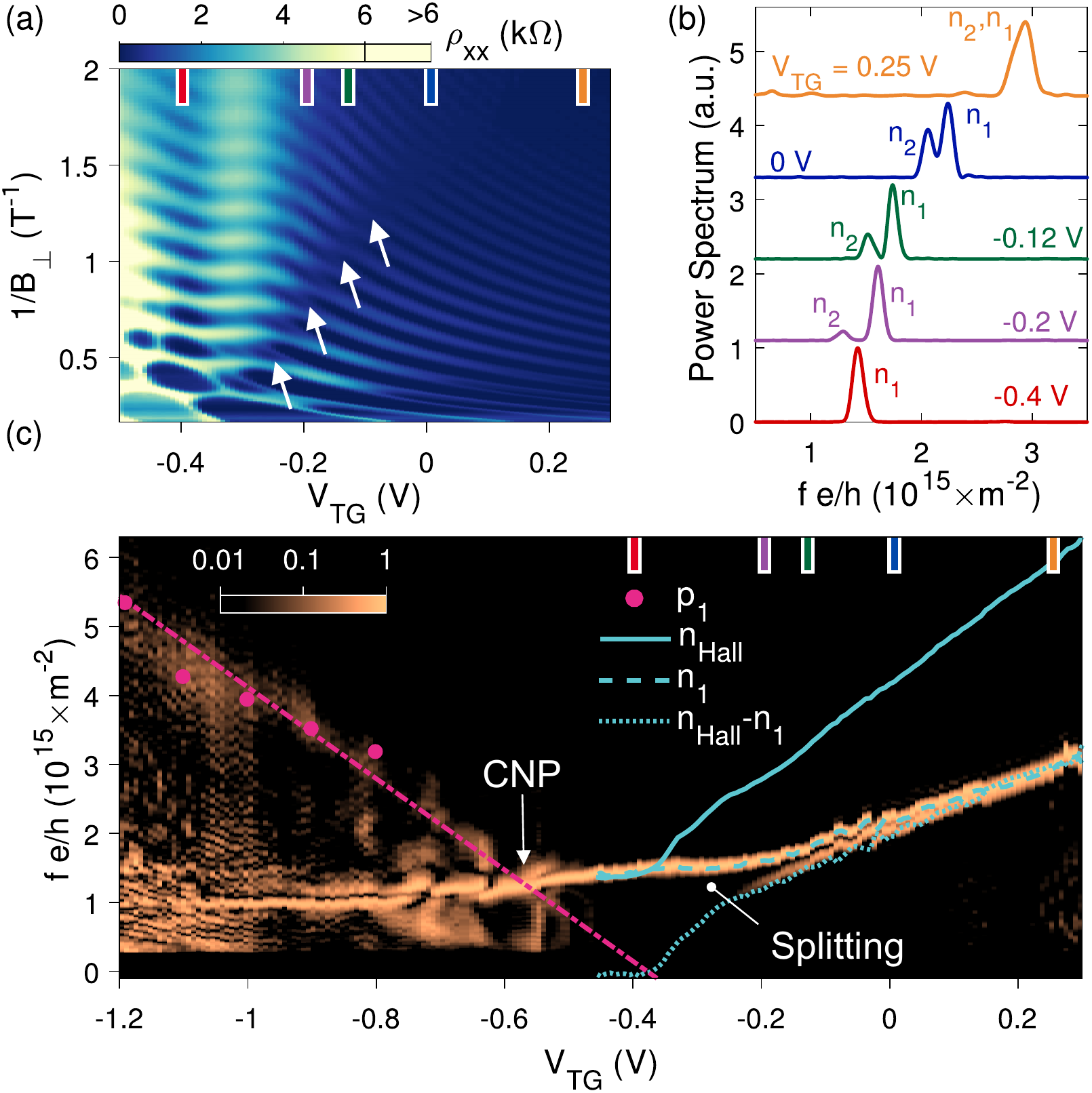}
\caption{(a) Longitudinal resistivity $\rho_{xx}$ as in Fig.~\ref{fig2}(c) for $V_{\mathrm{TG}}\geq-0.5~\mathrm{V}$ as a function of $1/B_{\perp}$. The arrows indicate a beating in the SdH oscillations, visible as a $\pi$ phase shift. (b) Normalized power spectrum of $\rho_{xx}(1/B_{\perp})$ for various gate voltages (data offset for clarity). The frequency axis has been multiplied by $e/h$ to directly show the subband densities. (c) Color map of the power spectrum as in (b) as a function of $V_{\mathrm{TG}}$. The amplitude of the power spectrum has been normalized, column by column, to the $n_1$ peak.
The solid blue line indicates the density obtained from the Hall slope, the dashed line marks the $n_1$ peak and the dotted line gives the difference between the two. Dots indicate the hole density obtained from hole-like LLs in Fig.~\ref{fig2}(c) with the dashed-dotted line being a guide to the eye.}
\label{fig3}
\end{figure}

Magnetotransport measurements, shown in Fig.~\ref{fig2}, confirm the sample has an inverted band structure, and is tunable from a pure electron regime to a mixed electron-hole regime. Typical for high mobility structures~\cite{Knez2010,Qu2015}, the longitudinal resistivity $\rho_{xx}$ exhibits a series of peaks and dips as a function of $V_{\mathrm{TG}}$, as shown in Fig.~\ref{fig2}(a). The resistance peaks at $V_{\mathrm{TG}}=-0.60~\mathrm{V}$ and $V_{\mathrm{TG}}=-0.35~\mathrm{V}$ are interpreted with the Fermi energy crossing the CNP and the valence band top respectively \cite{Qu2015}, as discussed in reference to Fig.~\ref{fig3}(c). In Ref.~\onlinecite{Knez2010} a resistance dip in the hole-dominated region, similar to what we observe at $V_{\mathrm{TG}}=-0.75~\mathrm{V}$, was identified as a van Hove singularity at the bottom of the hybridization gap.

Figure~\ref{fig2}(b) shows the transverse resistivity $\rho_{xy}$ as a function of perpendicular field $B_{\perp}$ for different values of $V_\mathrm{TG}$. For $V_\mathrm{TG}> -0.4~\mathrm{V}$, $\rho_{xy}$ has a positive slope, indicative of exclusively electron-like transport. For $V_\mathrm{TG}\leq -0.75~\mathrm{V}$, the $\rho_{xy}$ slope reverses at finite $B_\perp$, a hallmark of the simultaneous presence of electrons and holes in the system. This behavior persists down to $V_\mathrm{TG}=-1.2~\mathrm{V}$, indicating a pure hole state is not reached in the gate range of operation, consistent with the calculation of Fig.~\ref{fig1}(a).

The ambipolar behavior discussed above in terms of $\rho_{xy}$ also becomes apparent in $\rho_{xx}$ in large perpendicular magnetic fields, where Shubnikov-de Haas (SdH) oscillations and quantum Hall states develop in the entire gate range [Fig.~\ref{fig2}(c)]. For $V_{\mathrm{TG}}\geq -0.2~\mathrm{V}$ we observe regular electron-like Landau levels (LLs) with Zeeman splitting at high field, as indicated by the numbering in Fig.~\ref{fig2}(c), obtained from $\rho_{xy}$. The large resistance increase as a function of $B_{\perp}$ for $V_{\mathrm{TG}}\approx -0.6~\mathrm{V}$ is consistent with an identical number of electron and hole LLs at the CNP \cite{Nicholas2000,Nichele2014}.

For $V_{\mathrm{TG}}\leq -0.5~\mathrm{V}$ electron-like and hole-like LLs coexist, as also evident from the non-monotonic $\rho_{xy}$ [see Fig.~\ref{fig2}(b)]. In this regime, signatures of electron-hole hybridization are visible as avoided-crossings between LLs, as previously observed via cyclotron resonances~\cite{Tsay1997,Vasilyev1999}. Based on the analysis presented in Fig.~\ref{fig3}(c), we assign to the hole-like LLs the filling factors indicated with negative numbering.
Approaching the CNP from the electron regime, a peculiar closing and reopening of spin-split levels takes place, as marked with primed numbers. This is associated with the spin-orbit splitting becoming larger than the LL separation. An additional evolution of the LLs takes places for $V_{TG}\approx -0.4~\mathrm{V}$ as indicated with double-primed numbering. As discussed in the following, this is associated with the depopulation of one split subband. Filling factors assigned to primed and double primed LLs are confirmed by $\rho_{xy}$ measurements.

We now address the electron-like states close to the hybridization gap. Low-field SdH oscillations are a powerful tool to study properties at the Fermi surface such as electron density and effective mass ~\cite{Ando1982,Coleridge1989}. In systems where two subbands contribute to transport in parallel, as 2DEGs with strong SOC, the SdH oscillations manifest a beating pattern given by the superposition of two sets of oscillations with different $1/B_{\perp}$ periodicity~\cite{Bychkov1984,Heida1998,Luo1988,Das1989,Brosig1999,Gui2004,Nishioka2009}. The power spectrum of $\rho_{xx}~(1/B_{\perp})$ then allows one to extract the density components $n_{i}$ from the peak frequencies $f_{i}$ as $n_{i}=ef_{i}/h$~\cite{Winkler2003}. The SdH analysis gives the densities of the individual subbands and the Hall slope gives the net free charge of the system $n_{\mathrm{Hall}}$. For two spin-split electron-like subbands we expect $n_{\mathrm{Hall}}=n_{1}+n_{2}$.

Figure~\ref{fig3}(a) shows a zoom-in of Fig.~\ref{fig2}(c) for the electron regime with the vertical axis scaled as $1/B_{\perp}$ to make the SdH oscillations periodic. A beating, visible as a $\pi$ phase slip, is indicated with arrows. Figure~\ref{fig3}(b) shows the power spectrum of the data in Fig.~\ref{fig3}(a) for five gate voltage values. The frequency axis $f$ has been multiplied by $e/h$ to directly show the subband densities. At positive $V_{\mathrm{TG}}$, the power spectrum reveals a single oscillation frequency. Decreasing $V_{\mathrm{TG}}$, the peak moves to lower electron densities and gradually splits into two components. The amplitude of the low-density peak decreases with respect to its high density counterpart ($n_1$) until it disappears in the background for $V_{\mathrm{TG}}<-0.25~\mathrm{V}$. The quench of the $n_2$ peak at finite density is compatible with a $k\neq 0$ minimum in the dispersion relation of the high energy split band, as just above energy II in Fig.~\ref{fig1}(a).

\begin{figure}
\includegraphics[width=\columnwidth]{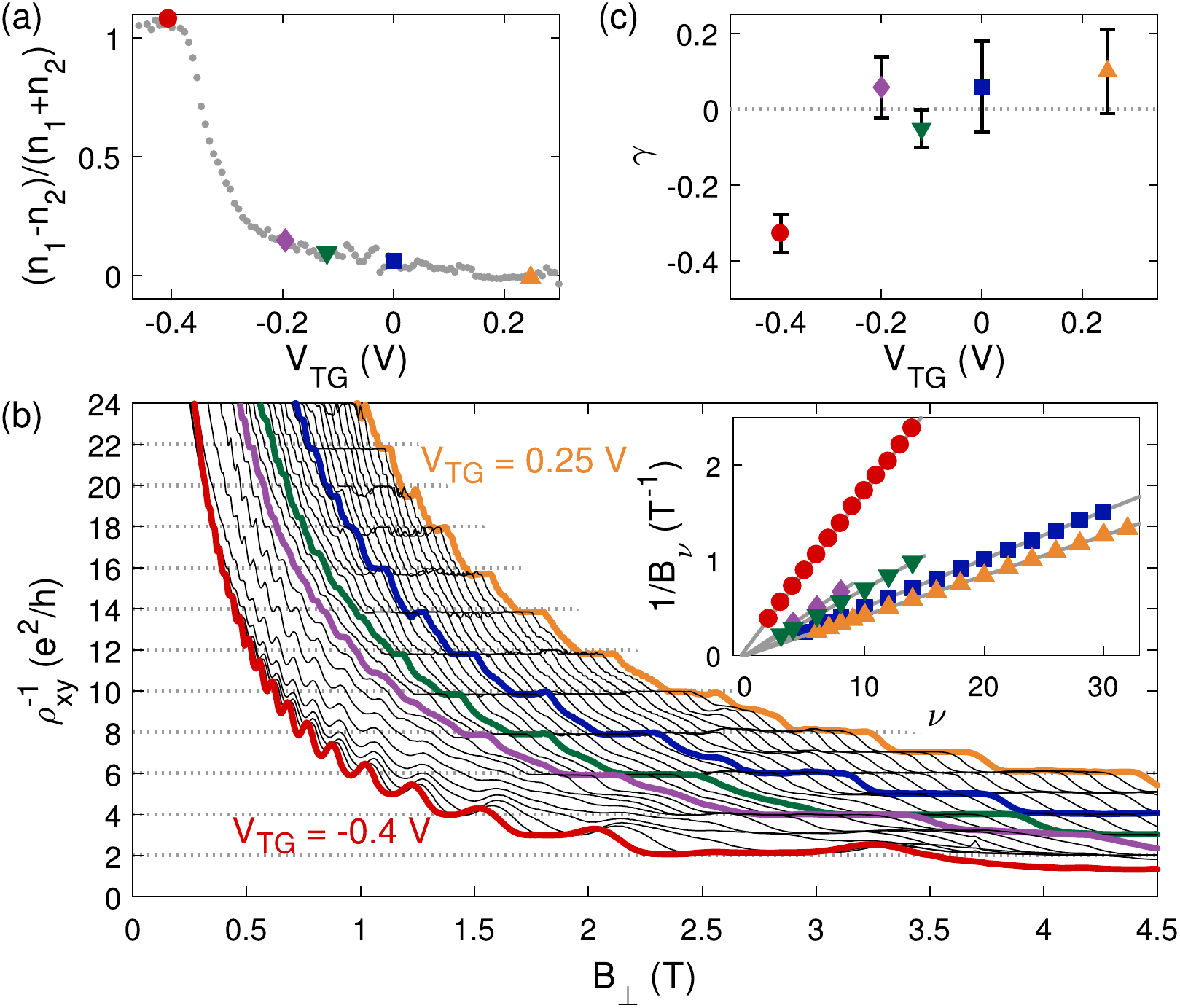}
\caption{(a) Spin-orbit polarization of electron-like states as a function of $V_{\mathrm{TG}}$, with markers defined as in (b). (b) Inverse transverse resistivity, $\rho_{xy}^{-1}$, for different top gate voltages, $V_{\mathrm{TG}}$. Inset: Inverse magnetic field positions of the filling factors $\nu$ for different $V_{\mathrm{TG}}$ values. Solid lines are linear fits to the data. (c) Phase offset $\gamma$ of the data in the inset of (b) extrapolated for $1/B\rightarrow 0$.}
\label{fig4}
\end{figure}

Additional insight into the data is gained by comparing the peak positions with the Hall density. The same analysis as in Fig.~\ref{fig3}(b) is shown in the color plot of Fig.~\ref{fig3}(c) as a function of $V_{\mathrm{TG}}$. The solid blue line indicates the density $n_{\mathrm{Hall}}$, extracted from $\rho_{xy}$. The dashed line tracks the position of the $n_1$ peak in the power spectrum while the dotted line shows the quantity $n_{\mathrm{Hall}}-n_1$. For $V_{\mathrm{TG}}>0$ a single peak is visible in the spectrum with $fe/h=n_{\mathrm{Hall}}/2$. This is consistent with two spin degenerate bands with $n_1=n_2$, as in scenario I in Fig.~\ref{fig1}(a). Once the splitting develops, as highlighted in Fig.~\ref{fig3}(c), $n_{\mathrm{Hall}}-n_1$ matches the position of the measured $n_2$ peak. 
The analysis is extended down to $V_{\mathrm{TG}}=-0.46~\mathrm{V}$, where $\rho_{xy}$ does not show indication of hole transport yet. The density difference between the two subbands gradually increases until $n_{\mathrm{Hall}}=n_1$ at $V_{\mathrm{TG}}\approx -0.4~\mathrm{V}$, i.e. all mobile charge resides in a \emph{single} band with $n_1=1.4\times 10^{15}~\mathrm{m^{-2}}$. This is compatible with situation II in Fig.~\ref{fig1}(a).

Below the CNP, the electron-like $n_1$ peak coexists with a hole-like state, highlighted with a dotted-dashed line in Fig.~\ref{fig3}(c). We confirmed that its position matches the periodicity of the hole-like LLs [cf. dots in Figs.~\ref{fig2}(c) and Figs.~\ref{fig3}(c)]. The hole signature in the spectrum can either be interpreted as two degenerate subbands $p_1=p_2$, or one spin-orbit polarized subband $p_1$. Extracting the total density from the Hall slope is less accurate in this regime due to the non-linearity of $\rho_{xy}(B)$, preventing further analysis. Nevertheless, assuming a single subband $p_1$, as predicted by our model for situation III in Fig.~\ref{fig1}(b), the top gate capacitance in the hole regime ($-\partial p_1/\partial V_{\mathrm{TG}}$) matches that in the electron regime ($\partial n_{\mathrm{Hall}}/\partial V_{\mathrm{TG}}$), as expected from the electrostatic model of Fig.~\ref{fig1}(a). Furthermore, the absence of Zeeman splitting in the hole-like LL up to high field supports the interpretation that holes are also fully spin-orbit polarized. Assuming a single hole-like band, the filling factors indicated in Fig.~\ref{fig3}(c) with negative numbering are calculated for the hole-like LLs, consistent with identical filling factor for electron- and hole-like LL ($1$ and $-1$ respectively) being populated at the CNP \cite{Nicholas2000,Nichele2014}. From these observation we conclude that a \emph{single} and \emph{fully spin-orbit polarized} hole band $p_1$ is occupied below the CNP, consistent with scenario III in Fig.~\ref{fig1}(b).

The intersection between $p_1$ and $n_1$ at $V_{\mathrm{TG}}\approx-0.6~\mathrm{V}$ determines the CNP, consistent with Fig.~\ref{fig2}(b). The crossing of the Fermi energy with the top of the valence band is inferred to be at $V_{\mathrm{TG}}=-0.35~\mathrm{V}$. This matches the peak in $\rho_{xx}$, as seen in Fig.~\ref{fig2}(a), and the kink in $n_{\mathrm{Hall}}$ visible in Fig.~\ref{fig3}(c) marking a change in gate capacitance as a screening layer is populated. 

After demonstrating the large splitting at the CNP, we investigate how the large spin-orbit polarization affects transport phenomena. The zero field polarization of electron-like states, quantified as $(n_1-n_2)/(n_1+n_2)$, saturates at $100\%$ for $V_{\mathrm{TG}}=-0.4~\mathrm{V}$ [Fig.~\ref{fig4}(a)]. Despite expecting hole-like states in this regime, hole conduction is not experimentally detected, either by a slope reversal in $\rho_{xy}$ [Fig.~\ref{fig2}(b)] or additional LLs in $\rho_{xx}$ [Fig.~\ref{fig2}(c)]. This behavior is presumably due to the low mobility of holes in GaSb which, for densities lower than $5\times 10^{14}~\mathrm{m^{-2}}$ may localize. As only electron-like states contribute to transport, this situation effectively realizes a helical 2DEG. Such a system is reminiscent of the surface of three-dimensional topological insulators, where the Fermi energy crosses a single spin resolved band, and might have potential interest for studying topological states of matter.

The full spin-orbit polarization for $V_{\mathrm{TG}}\approx-0.4~\mathrm{V}$ is further confirmed by the quantum Hall plateaus of $\rho_{xy}^{-1}$, shown in Fig.~\ref{fig4}(b). At high electron density (orange line, $V_{\mathrm{TG}}=0.25~\mathrm{V}$) the plateaus evolve in steps of $2e^2/h$, as expected for a conventional 2DEG. For $B_\perp>3~\mathrm{T}$, Zeeman splitting lifts spin degeneracy resulting in $e^2/h$ plateaus. In the fully polarized regime (red line, $V_{\mathrm{TG}}=-0.4~\mathrm{V}$) the plateaus exquisitely evolve as integer multiples of $e^2/h$ from the first visible steps at $B_{\perp}\approx 400~\mathrm{mT}$. This is further evidence of the helical nature of electron-like states, extending also to small magnetic fields. The oscillations in the low density plateaus [also visible in Fig.~\ref{fig2}(b)] are attributed to disorder, resulting in a broadening of LLs and an eventual mixing between $\rho_{xx}$ and $\rho_{xy}$ \cite{Huckestein1995}. We note that the overshoots in $\rho_{xy}^{-1}$ or an eventual presence of hole-like states do not compromise the analysis. In fact the density of the system for $V_{\mathrm{TG}}=-0.4~\mathrm{V}$ is confirmed within $5\%$ by three independent checks: (i) The slope of $\rho_{xy}$, constant up to $B_{\perp}=5~\mathrm{T}$, (ii) The periodicity of the low-field SdH oscillations, (iii) The magnetic field position $B_\nu$ of the $\nu e^2/h$ plateaus in $\rho_{xy}^{-1}$, satisfying $n_1=\nu e B_\nu/h$.

The unique Fermi level crossing present in our system, together with strong SOC, can result in a non-trivial Berry phase acquired by electrons on a closed cyclotron orbit, such as in Fig.~\ref{fig1}(c). To check this eventuality, we measured the phase offset $\varphi$ of the SdH oscillations for $1/B\rightarrow 0$, similar to earlier work on graphene \cite{Novoselov2005,Zhang2005} and 3D topological insulators \cite{Qu2010,Xiong2012}. While conventional 2DEGs have $\varphi=0$, materials with a symmetric Dirac cone exhibit $\varphi=1/2$. In a complex band structure as in the present case, the Berry phase is not expected to be quantized but to vary depending on the details of the dispersion relation \cite{Wright2013}.
The inset of Figure~\ref{fig4}(b) shows the $1/B_\nu$ positions of the $\nu$ filling factors for various top gate voltages (markers) together with linear fits (lines) extrapolating to $1/B\rightarrow 0$. The result of the extrapolation is shown in Fig.~\ref{fig4}(c). For $V_{\mathrm{TG}}\geq-0.2~\mathrm{V}$, all the curves consistently give $\varphi\approx 0$, as expected for normal fermions. For $V_{\mathrm{TG}}=-0.4~\mathrm{V}$ the extrapolation leads a phase shift $\varphi=-0.33\pm 0.05$, consistent with a non-zero Berry phase.

In conclusion, we studied the band structure of inverted InAs/GaSb QWs via magnetotransport measurements. Consistent with simulations, electron-like and hole-like states are fully spin-orbit polarized in proximity of the CNP. We identify a regime where a single electron-like band with helical spin texture contributes to transport. The $100\%$ spin-orbit polarization of the system is confirmed by quantum Hall plateaus evolving in $e^2/h$ steps and a non-trivial Berry phase.

\begin{acknowledgments}
This work was supported from Microsoft Corporation Station Q. The work in Copenhagen was also supported from the Danish National Research Foundation and the Villum Foundation. The work in Delft was also supported by the Dutch Organisation  for  Scientific Research (NWO) and the Foundation for Fundamental Research on Matter (FOM).
We thank Emmanuel Rashba, Joshua Folk and Karsten Flensberg for valuable discussions. F.N. acknowledges support of the European Commission through the Marie Curie Fellowship, grant agreement No 659653.
\end{acknowledgments}

\bibliography{Bibliography}

\begin{thebibliography}{53}%
\makeatletter
\providecommand \@ifxundefined [1]{%
 \@ifx{#1\undefined}
}%
\providecommand \@ifnum [1]{%
 \ifnum #1\expandafter \@firstoftwo
 \else \expandafter \@secondoftwo
 \fi
}%
\providecommand \@ifx [1]{%
 \ifx #1\expandafter \@firstoftwo
 \else \expandafter \@secondoftwo
 \fi
}%
\providecommand \natexlab [1]{#1}%
\providecommand \enquote  [1]{``#1''}%
\providecommand \bibnamefont  [1]{#1}%
\providecommand \bibfnamefont [1]{#1}%
\providecommand \citenamefont [1]{#1}%
\providecommand \href@noop [0]{\@secondoftwo}%
\providecommand \href [0]{\begingroup \@sanitize@url \@href}%
\providecommand \@href[1]{\@@startlink{#1}\@@href}%
\providecommand \@@href[1]{\endgroup#1\@@endlink}%
\providecommand \@sanitize@url [0]{\catcode `\\12\catcode `\$12\catcode
  `\&12\catcode `\#12\catcode `\^12\catcode `\_12\catcode `\%12\relax}%
\providecommand \@@startlink[1]{}%
\providecommand \@@endlink[0]{}%
\providecommand \url  [0]{\begingroup\@sanitize@url \@url }%
\providecommand \@url [1]{\endgroup\@href {#1}{\urlprefix }}%
\providecommand \urlprefix  [0]{URL }%
\providecommand \Eprint [0]{\href }%
\providecommand \doibase [0]{http://dx.doi.org/}%
\providecommand \selectlanguage [0]{\@gobble}%
\providecommand \bibinfo  [0]{\@secondoftwo}%
\providecommand \bibfield  [0]{\@secondoftwo}%
\providecommand \translation [1]{[#1]}%
\providecommand \BibitemOpen [0]{}%
\providecommand \bibitemStop [0]{}%
\providecommand \bibitemNoStop [0]{.\EOS\space}%
\providecommand \EOS [0]{\spacefactor3000\relax}%
\providecommand \BibitemShut  [1]{\csname bibitem#1\endcsname}%
\let\auto@bib@innerbib\@empty
\bibitem [{\citenamefont {Yang}\ \emph {et~al.}(1997)\citenamefont {Yang},
  \citenamefont {Yang}, \citenamefont {Bennett},\ and\ \citenamefont
  {Shanabrook}}]{Yang1997}%
  \BibitemOpen
  \bibfield  {author} {\bibinfo {author} {\bibfnamefont {M.~J.}\ \bibnamefont
  {Yang}}, \bibinfo {author} {\bibfnamefont {C.~H.}\ \bibnamefont {Yang}},
  \bibinfo {author} {\bibfnamefont {B.~R.}\ \bibnamefont {Bennett}}, \ and\
  \bibinfo {author} {\bibfnamefont {B.~V.}\ \bibnamefont {Shanabrook}},\ }\href
  {http://link.aps.org/doi/10.1103/PhysRevLett.78.4613} {\bibfield  {journal}
  {\bibinfo  {journal} {Phys. Rev. Lett.}\ }\textbf {\bibinfo {volume} {78}},\
  \bibinfo {pages} {4613} (\bibinfo {year} {1997})}\BibitemShut {NoStop}%
\bibitem [{\citenamefont {Liu}\ \emph {et~al.}(2008)\citenamefont {Liu},
  \citenamefont {Hughes}, \citenamefont {Qi}, \citenamefont {Wang},\ and\
  \citenamefont {Zhang}}]{Liu2008}%
  \BibitemOpen
  \bibfield  {author} {\bibinfo {author} {\bibfnamefont {C.}~\bibnamefont
  {Liu}}, \bibinfo {author} {\bibfnamefont {T.~L.}\ \bibnamefont {Hughes}},
  \bibinfo {author} {\bibfnamefont {X.-L.}\ \bibnamefont {Qi}}, \bibinfo
  {author} {\bibfnamefont {K.}~\bibnamefont {Wang}}, \ and\ \bibinfo {author}
  {\bibfnamefont {S.-C.}\ \bibnamefont {Zhang}},\ }\href
  {http://link.aps.org/doi/10.1103/PhysRevLett.100.236601} {\bibfield
  {journal} {\bibinfo  {journal} {Phys. Rev. Lett.}\ }\textbf {\bibinfo
  {volume} {100}},\ \bibinfo {pages} {236601} (\bibinfo {year}
  {2008})}\BibitemShut {NoStop}%
\bibitem [{\citenamefont {Qu}\ \emph {et~al.}(2015)\citenamefont {Qu},
  \citenamefont {Beukman}, \citenamefont {Nadj-Perge}, \citenamefont {Wimmer},
  \citenamefont {Nguyen}, \citenamefont {Yi}, \citenamefont {Thorp},
  \citenamefont {Sokolich}, \citenamefont {Kiselev}, \citenamefont {Manfra},
  \citenamefont {Marcus},\ and\ \citenamefont {Kouwenhoven}}]{Qu2015}%
  \BibitemOpen
  \bibfield  {author} {\bibinfo {author} {\bibfnamefont {F.}~\bibnamefont
  {Qu}}, \bibinfo {author} {\bibfnamefont {A.~J.~A.}\ \bibnamefont {Beukman}},
  \bibinfo {author} {\bibfnamefont {S.}~\bibnamefont {Nadj-Perge}}, \bibinfo
  {author} {\bibfnamefont {M.}~\bibnamefont {Wimmer}}, \bibinfo {author}
  {\bibfnamefont {B.-M.}\ \bibnamefont {Nguyen}}, \bibinfo {author}
  {\bibfnamefont {W.}~\bibnamefont {Yi}}, \bibinfo {author} {\bibfnamefont
  {J.}~\bibnamefont {Thorp}}, \bibinfo {author} {\bibfnamefont
  {M.}~\bibnamefont {Sokolich}}, \bibinfo {author} {\bibfnamefont {A.~A.}\
  \bibnamefont {Kiselev}}, \bibinfo {author} {\bibfnamefont {M.~J.}\
  \bibnamefont {Manfra}}, \bibinfo {author} {\bibfnamefont {C.~M.}\
  \bibnamefont {Marcus}}, \ and\ \bibinfo {author} {\bibfnamefont {L.~P.}\
  \bibnamefont {Kouwenhoven}},\ }\href
  {http://link.aps.org/doi/10.1103/PhysRevLett.115.036803} {\bibfield
  {journal} {\bibinfo  {journal} {Phys. Rev. Lett.}\ }\textbf {\bibinfo
  {volume} {115}},\ \bibinfo {pages} {036803} (\bibinfo {year}
  {2015})}\BibitemShut {NoStop}%
\bibitem [{\citenamefont {{Nichele}}\ \emph {et~al.}(2015)\citenamefont
  {{Nichele}}, \citenamefont {{Suominen}}, \citenamefont {{Kjaergaard}},
  \citenamefont {{Marcus}}, \citenamefont {{Sajadi}}, \citenamefont {{Folk}},
  \citenamefont {{Qu}}, \citenamefont {{Beukman}}, \citenamefont {{de Vries}},
  \citenamefont {{van Veen}}, \citenamefont {{Nadj-Perge}}, \citenamefont
  {{Kouwenhoven}}, \citenamefont {{Nguyen}}, \citenamefont {{Kiselev}},
  \citenamefont {{Yi}}, \citenamefont {{Sokolich}}, \citenamefont {{Manfra}},
  \citenamefont {{Spanton}},\ and\ \citenamefont {{Moler}}}]{Nichele2015}%
  \BibitemOpen
  \bibfield  {author} {\bibinfo {author} {\bibfnamefont {F.}~\bibnamefont
  {{Nichele}}}, \bibinfo {author} {\bibfnamefont {H.~J.}\ \bibnamefont
  {{Suominen}}}, \bibinfo {author} {\bibfnamefont {M.}~\bibnamefont
  {{Kjaergaard}}}, \bibinfo {author} {\bibfnamefont {C.~M.}\ \bibnamefont
  {{Marcus}}}, \bibinfo {author} {\bibfnamefont {E.}~\bibnamefont {{Sajadi}}},
  \bibinfo {author} {\bibfnamefont {J.~A.}\ \bibnamefont {{Folk}}}, \bibinfo
  {author} {\bibfnamefont {F.}~\bibnamefont {{Qu}}}, \bibinfo {author}
  {\bibfnamefont {A.~J.~A.}\ \bibnamefont {{Beukman}}}, \bibinfo {author}
  {\bibfnamefont {F.~K.}\ \bibnamefont {{de Vries}}}, \bibinfo {author}
  {\bibfnamefont {J.}~\bibnamefont {{van Veen}}}, \bibinfo {author}
  {\bibfnamefont {S.}~\bibnamefont {{Nadj-Perge}}}, \bibinfo {author}
  {\bibfnamefont {L.~P.}\ \bibnamefont {{Kouwenhoven}}}, \bibinfo {author}
  {\bibfnamefont {B.-M.}\ \bibnamefont {{Nguyen}}}, \bibinfo {author}
  {\bibfnamefont {A.~A.}\ \bibnamefont {{Kiselev}}}, \bibinfo {author}
  {\bibfnamefont {W.}~\bibnamefont {{Yi}}}, \bibinfo {author} {\bibfnamefont
  {M.}~\bibnamefont {{Sokolich}}}, \bibinfo {author} {\bibfnamefont {M.~J.}\
  \bibnamefont {{Manfra}}}, \bibinfo {author} {\bibfnamefont {E.~M.}\
  \bibnamefont {{Spanton}}}, \ and\ \bibinfo {author} {\bibfnamefont {K.~A.}\
  \bibnamefont {{Moler}}},\ }\href@noop {} {\bibfield  {journal} {\bibinfo
  {journal} {ArXiv e-prints}\ } (\bibinfo {year} {2015})},\ \Eprint
  {http://arxiv.org/abs/1511.01728} {arXiv:1511.01728 [cond-mat.mes-hall]}
  \BibitemShut {NoStop}%
\bibitem [{\citenamefont {Knez}\ \emph {et~al.}(2011)\citenamefont {Knez},
  \citenamefont {Du},\ and\ \citenamefont {Sullivan}}]{Knez2011}%
  \BibitemOpen
  \bibfield  {author} {\bibinfo {author} {\bibfnamefont {I.}~\bibnamefont
  {Knez}}, \bibinfo {author} {\bibfnamefont {R.-R.}\ \bibnamefont {Du}}, \ and\
  \bibinfo {author} {\bibfnamefont {G.}~\bibnamefont {Sullivan}},\ }\href
  {http://link.aps.org/doi/10.1103/PhysRevLett.107.136603} {\bibfield
  {journal} {\bibinfo  {journal} {Phys. Rev. Lett.}\ }\textbf {\bibinfo
  {volume} {107}},\ \bibinfo {pages} {136603} (\bibinfo {year}
  {2011})}\BibitemShut {NoStop}%
\bibitem [{\citenamefont {Suzuki}\ \emph {et~al.}(2013)\citenamefont {Suzuki},
  \citenamefont {Harada}, \citenamefont {Onomitsu},\ and\ \citenamefont
  {Muraki}}]{Suzuki2013}%
  \BibitemOpen
  \bibfield  {author} {\bibinfo {author} {\bibfnamefont {K.}~\bibnamefont
  {Suzuki}}, \bibinfo {author} {\bibfnamefont {Y.}~\bibnamefont {Harada}},
  \bibinfo {author} {\bibfnamefont {K.}~\bibnamefont {Onomitsu}}, \ and\
  \bibinfo {author} {\bibfnamefont {K.}~\bibnamefont {Muraki}},\ }\href
  {http://link.aps.org/doi/10.1103/PhysRevB.87.235311} {\bibfield  {journal}
  {\bibinfo  {journal} {Phys. Rev. B}\ }\textbf {\bibinfo {volume} {87}},\
  \bibinfo {pages} {235311} (\bibinfo {year} {2013})}\BibitemShut {NoStop}%
\bibitem [{\citenamefont {Knez}\ \emph {et~al.}(2014)\citenamefont {Knez},
  \citenamefont {Rettner}, \citenamefont {Yang}, \citenamefont {Parkin},
  \citenamefont {Du}, \citenamefont {Du},\ and\ \citenamefont
  {Sullivan}}]{Knez2014}%
  \BibitemOpen
  \bibfield  {author} {\bibinfo {author} {\bibfnamefont {I.}~\bibnamefont
  {Knez}}, \bibinfo {author} {\bibfnamefont {C.~T.}\ \bibnamefont {Rettner}},
  \bibinfo {author} {\bibfnamefont {S.-H.}\ \bibnamefont {Yang}}, \bibinfo
  {author} {\bibfnamefont {S.~S.~P.}\ \bibnamefont {Parkin}}, \bibinfo {author}
  {\bibfnamefont {L.}~\bibnamefont {Du}}, \bibinfo {author} {\bibfnamefont
  {R.-R.}\ \bibnamefont {Du}}, \ and\ \bibinfo {author} {\bibfnamefont
  {G.}~\bibnamefont {Sullivan}},\ }\href {\doibase
  10.1103/PhysRevLett.112.026602} {\bibfield  {journal} {\bibinfo  {journal}
  {Phys. Rev. Lett.}\ }\textbf {\bibinfo {volume} {112}},\ \bibinfo {pages}
  {026602} (\bibinfo {year} {2014})}\BibitemShut {NoStop}%
\bibitem [{\citenamefont {Du}\ \emph {et~al.}(2015)\citenamefont {Du},
  \citenamefont {Knez}, \citenamefont {Sullivan},\ and\ \citenamefont
  {Du}}]{Du2015}%
  \BibitemOpen
  \bibfield  {author} {\bibinfo {author} {\bibfnamefont {L.}~\bibnamefont
  {Du}}, \bibinfo {author} {\bibfnamefont {I.}~\bibnamefont {Knez}}, \bibinfo
  {author} {\bibfnamefont {G.}~\bibnamefont {Sullivan}}, \ and\ \bibinfo
  {author} {\bibfnamefont {R.-R.}\ \bibnamefont {Du}},\ }\href {\doibase
  10.1103/PhysRevLett.114.096802} {\bibfield  {journal} {\bibinfo  {journal}
  {Phys. Rev. Lett.}\ }\textbf {\bibinfo {volume} {114}},\ \bibinfo {pages}
  {096802} (\bibinfo {year} {2015})}\BibitemShut {NoStop}%
\bibitem [{\citenamefont {Mueller}\ \emph {et~al.}(2015)\citenamefont
  {Mueller}, \citenamefont {Pal}, \citenamefont {Karalic}, \citenamefont
  {Tschirky}, \citenamefont {Charpentier}, \citenamefont {Wegscheider},
  \citenamefont {Ensslin},\ and\ \citenamefont {Ihn}}]{Mueller2015}%
  \BibitemOpen
  \bibfield  {author} {\bibinfo {author} {\bibfnamefont {S.}~\bibnamefont
  {Mueller}}, \bibinfo {author} {\bibfnamefont {A.~N.}\ \bibnamefont {Pal}},
  \bibinfo {author} {\bibfnamefont {M.}~\bibnamefont {Karalic}}, \bibinfo
  {author} {\bibfnamefont {T.}~\bibnamefont {Tschirky}}, \bibinfo {author}
  {\bibfnamefont {C.}~\bibnamefont {Charpentier}}, \bibinfo {author}
  {\bibfnamefont {W.}~\bibnamefont {Wegscheider}}, \bibinfo {author}
  {\bibfnamefont {K.}~\bibnamefont {Ensslin}}, \ and\ \bibinfo {author}
  {\bibfnamefont {T.}~\bibnamefont {Ihn}},\ }\href
  {http://link.aps.org/doi/10.1103/PhysRevB.92.081303} {\bibfield  {journal}
  {\bibinfo  {journal} {Phys. Rev. B}\ }\textbf {\bibinfo {volume} {92}},\
  \bibinfo {pages} {081303} (\bibinfo {year} {2015})}\BibitemShut {NoStop}%
\bibitem [{\citenamefont {\v{Z}uti\'c Igor}\ \emph {et~al.}(2004)\citenamefont
  {\v{Z}uti\'c Igor}, \citenamefont {Fabian},\ and\ \citenamefont
  {Das~Sarma}}]{Zutic2004}%
  \BibitemOpen
  \bibfield  {author} {\bibinfo {author} {\bibnamefont {\v{Z}uti\'c Igor}},
  \bibinfo {author} {\bibfnamefont {J.}~\bibnamefont {Fabian}}, \ and\ \bibinfo
  {author} {\bibfnamefont {S.}~\bibnamefont {Das~Sarma}},\ }\href
  {http://link.aps.org/doi/10.1103/RevModPhys.76.323} {\bibfield  {journal}
  {\bibinfo  {journal} {Rev. Mod. Phys.}\ }\textbf {\bibinfo {volume} {76}},\
  \bibinfo {pages} {323} (\bibinfo {year} {2004})}\BibitemShut {NoStop}%
\bibitem [{Not()}]{Note_winkler}%
  \BibitemOpen
  \href@noop {} {}\bibinfo {note} {We use the same subband definition as in
  Sec. 6.3.1 of Ref.~\cite{Winkler2003}.}\BibitemShut {Stop}%
\bibitem [{\citenamefont {Winkler}(2003)}]{Winkler2003}%
  \BibitemOpen
  \bibfield  {author} {\bibinfo {author} {\bibfnamefont {R.}~\bibnamefont
  {Winkler}},\ }\href@noop {} {\emph {\bibinfo {title} {Spin-Orbit Coupling
  Effects in Two-Dimensional Electron and Hole Systems}}},\ \bibinfo {series}
  {Springer Tracts in Modern Physics}, Vol.\ \bibinfo {volume} {191}\ (\bibinfo
   {publisher} {Springer-Verlag, Berlin},\ \bibinfo {year} {2003})\BibitemShut
  {NoStop}%
\bibitem [{\citenamefont {Luo}\ \emph {et~al.}(1988)\citenamefont {Luo},
  \citenamefont {Munekata}, \citenamefont {Fang},\ and\ \citenamefont
  {Stiles}}]{Luo1988}%
  \BibitemOpen
  \bibfield  {author} {\bibinfo {author} {\bibfnamefont {J.}~\bibnamefont
  {Luo}}, \bibinfo {author} {\bibfnamefont {H.}~\bibnamefont {Munekata}},
  \bibinfo {author} {\bibfnamefont {F.~F.}\ \bibnamefont {Fang}}, \ and\
  \bibinfo {author} {\bibfnamefont {P.~J.}\ \bibnamefont {Stiles}},\ }\href
  {\doibase 10.1103/PhysRevB.38.10142} {\bibfield  {journal} {\bibinfo
  {journal} {Phys. Rev. B}\ }\textbf {\bibinfo {volume} {38}},\ \bibinfo
  {pages} {10142} (\bibinfo {year} {1988})}\BibitemShut {NoStop}%
\bibitem [{\citenamefont {Das}\ \emph {et~al.}(1989)\citenamefont {Das},
  \citenamefont {Miller}, \citenamefont {Datta}, \citenamefont {Reifenberger},
  \citenamefont {Hong}, \citenamefont {Bhattacharya}, \citenamefont {Singh},\
  and\ \citenamefont {Jaffe}}]{Das1989}%
  \BibitemOpen
  \bibfield  {author} {\bibinfo {author} {\bibfnamefont {B.}~\bibnamefont
  {Das}}, \bibinfo {author} {\bibfnamefont {D.~C.}\ \bibnamefont {Miller}},
  \bibinfo {author} {\bibfnamefont {S.}~\bibnamefont {Datta}}, \bibinfo
  {author} {\bibfnamefont {R.}~\bibnamefont {Reifenberger}}, \bibinfo {author}
  {\bibfnamefont {W.~P.}\ \bibnamefont {Hong}}, \bibinfo {author}
  {\bibfnamefont {P.~K.}\ \bibnamefont {Bhattacharya}}, \bibinfo {author}
  {\bibfnamefont {J.}~\bibnamefont {Singh}}, \ and\ \bibinfo {author}
  {\bibfnamefont {M.}~\bibnamefont {Jaffe}},\ }\href {\doibase
  10.1103/PhysRevB.39.1411} {\bibfield  {journal} {\bibinfo  {journal} {Phys.
  Rev. B}\ }\textbf {\bibinfo {volume} {39}},\ \bibinfo {pages} {1411}
  (\bibinfo {year} {1989})}\BibitemShut {NoStop}%
\bibitem [{\citenamefont {Heida}\ \emph {et~al.}(1998)\citenamefont {Heida},
  \citenamefont {van Wees}, \citenamefont {Kuipers}, \citenamefont {Klapwijk},\
  and\ \citenamefont {Borghs}}]{Heida1998}%
  \BibitemOpen
  \bibfield  {author} {\bibinfo {author} {\bibfnamefont {J.~P.}\ \bibnamefont
  {Heida}}, \bibinfo {author} {\bibfnamefont {B.~J.}\ \bibnamefont {van Wees}},
  \bibinfo {author} {\bibfnamefont {J.~J.}\ \bibnamefont {Kuipers}}, \bibinfo
  {author} {\bibfnamefont {T.~M.}\ \bibnamefont {Klapwijk}}, \ and\ \bibinfo
  {author} {\bibfnamefont {G.}~\bibnamefont {Borghs}},\ }\href
  {http://link.aps.org/doi/10.1103/PhysRevB.57.11911} {\bibfield  {journal}
  {\bibinfo  {journal} {Phys. Rev. B}\ }\textbf {\bibinfo {volume} {57}},\
  \bibinfo {pages} {11911} (\bibinfo {year} {1998})}\BibitemShut {NoStop}%
\bibitem [{\citenamefont {Brosig}\ \emph {et~al.}(1999)\citenamefont {Brosig},
  \citenamefont {Ensslin}, \citenamefont {Warburton}, \citenamefont {Nguyen},
  \citenamefont {Brar}, \citenamefont {Thomas},\ and\ \citenamefont
  {Kroemer}}]{Brosig1999}%
  \BibitemOpen
  \bibfield  {author} {\bibinfo {author} {\bibfnamefont {S.}~\bibnamefont
  {Brosig}}, \bibinfo {author} {\bibfnamefont {K.}~\bibnamefont {Ensslin}},
  \bibinfo {author} {\bibfnamefont {R.~J.}\ \bibnamefont {Warburton}}, \bibinfo
  {author} {\bibfnamefont {C.}~\bibnamefont {Nguyen}}, \bibinfo {author}
  {\bibfnamefont {B.}~\bibnamefont {Brar}}, \bibinfo {author} {\bibfnamefont
  {M.}~\bibnamefont {Thomas}}, \ and\ \bibinfo {author} {\bibfnamefont
  {H.}~\bibnamefont {Kroemer}},\ }\href
  {http://link.aps.org/doi/10.1103/PhysRevB.60.R13989} {\bibfield  {journal}
  {\bibinfo  {journal} {Phys. Rev. B}\ }\textbf {\bibinfo {volume} {60}},\
  \bibinfo {pages} {R13989} (\bibinfo {year} {1999})}\BibitemShut {NoStop}%
\bibitem [{\citenamefont {Gui}\ \emph {et~al.}(2004)\citenamefont {Gui},
  \citenamefont {Becker}, \citenamefont {Dai}, \citenamefont {Liu},
  \citenamefont {Qiu}, \citenamefont {Novik}, \citenamefont {Sch\"afer},
  \citenamefont {Shu}, \citenamefont {Chu}, \citenamefont {Buhmann},\ and\
  \citenamefont {Molenkamp}}]{Gui2004}%
  \BibitemOpen
  \bibfield  {author} {\bibinfo {author} {\bibfnamefont {Y.~S.}\ \bibnamefont
  {Gui}}, \bibinfo {author} {\bibfnamefont {C.~R.}\ \bibnamefont {Becker}},
  \bibinfo {author} {\bibfnamefont {N.}~\bibnamefont {Dai}}, \bibinfo {author}
  {\bibfnamefont {J.}~\bibnamefont {Liu}}, \bibinfo {author} {\bibfnamefont
  {Z.~J.}\ \bibnamefont {Qiu}}, \bibinfo {author} {\bibfnamefont {E.~G.}\
  \bibnamefont {Novik}}, \bibinfo {author} {\bibfnamefont {M.}~\bibnamefont
  {Sch\"afer}}, \bibinfo {author} {\bibfnamefont {X.~Z.}\ \bibnamefont {Shu}},
  \bibinfo {author} {\bibfnamefont {J.~H.}\ \bibnamefont {Chu}}, \bibinfo
  {author} {\bibfnamefont {H.}~\bibnamefont {Buhmann}}, \ and\ \bibinfo
  {author} {\bibfnamefont {L.~W.}\ \bibnamefont {Molenkamp}},\ }\href {\doibase
  10.1103/PhysRevB.70.115328} {\bibfield  {journal} {\bibinfo  {journal} {Phys.
  Rev. B}\ }\textbf {\bibinfo {volume} {70}},\ \bibinfo {pages} {115328}
  (\bibinfo {year} {2004})}\BibitemShut {NoStop}%
\bibitem [{\citenamefont {Nishioka}\ \emph {et~al.}(2009)\citenamefont
  {Nishioka}, \citenamefont {Gurney}, \citenamefont {Marinero},\ and\
  \citenamefont {Mireles}}]{Nishioka2009}%
  \BibitemOpen
  \bibfield  {author} {\bibinfo {author} {\bibfnamefont {M.}~\bibnamefont
  {Nishioka}}, \bibinfo {author} {\bibfnamefont {B.~A.}\ \bibnamefont
  {Gurney}}, \bibinfo {author} {\bibfnamefont {E.~E.}\ \bibnamefont
  {Marinero}}, \ and\ \bibinfo {author} {\bibfnamefont {F.}~\bibnamefont
  {Mireles}},\ }\href {\doibase http://dx.doi.org/10.1063/1.3274129} {\bibfield
   {journal} {\bibinfo  {journal} {Applied Physics Letters}\ }\textbf {\bibinfo
  {volume} {95}},\ \bibinfo {pages} {242108} (\bibinfo {year}
  {2009})}\BibitemShut {NoStop}%
\bibitem [{\citenamefont {Stormer}\ \emph {et~al.}(1983)\citenamefont
  {Stormer}, \citenamefont {Schlesinger}, \citenamefont {Chang}, \citenamefont
  {Tsui}, \citenamefont {Gossard},\ and\ \citenamefont
  {Wiegmann}}]{Stormer1983}%
  \BibitemOpen
  \bibfield  {author} {\bibinfo {author} {\bibfnamefont {H.~L.}\ \bibnamefont
  {Stormer}}, \bibinfo {author} {\bibfnamefont {Z.}~\bibnamefont
  {Schlesinger}}, \bibinfo {author} {\bibfnamefont {A.}~\bibnamefont {Chang}},
  \bibinfo {author} {\bibfnamefont {D.~C.}\ \bibnamefont {Tsui}}, \bibinfo
  {author} {\bibfnamefont {A.~C.}\ \bibnamefont {Gossard}}, \ and\ \bibinfo
  {author} {\bibfnamefont {W.}~\bibnamefont {Wiegmann}},\ }\href {\doibase
  10.1103/PhysRevLett.51.126} {\bibfield  {journal} {\bibinfo  {journal} {Phys.
  Rev. Lett.}\ }\textbf {\bibinfo {volume} {51}},\ \bibinfo {pages} {126}
  (\bibinfo {year} {1983})}\BibitemShut {NoStop}%
\bibitem [{\citenamefont {Habib}\ \emph {et~al.}(2004)\citenamefont {Habib},
  \citenamefont {Tutuc}, \citenamefont {Melinte}, \citenamefont {Shayegan},
  \citenamefont {Wasserman}, \citenamefont {Lyon},\ and\ \citenamefont
  {Winkler}}]{Habib2004}%
  \BibitemOpen
  \bibfield  {author} {\bibinfo {author} {\bibfnamefont {B.}~\bibnamefont
  {Habib}}, \bibinfo {author} {\bibfnamefont {E.}~\bibnamefont {Tutuc}},
  \bibinfo {author} {\bibfnamefont {S.}~\bibnamefont {Melinte}}, \bibinfo
  {author} {\bibfnamefont {M.}~\bibnamefont {Shayegan}}, \bibinfo {author}
  {\bibfnamefont {D.}~\bibnamefont {Wasserman}}, \bibinfo {author}
  {\bibfnamefont {S.~A.}\ \bibnamefont {Lyon}}, \ and\ \bibinfo {author}
  {\bibfnamefont {R.}~\bibnamefont {Winkler}},\ }\href {\doibase
  10.1103/PhysRevB.69.113311} {\bibfield  {journal} {\bibinfo  {journal} {Phys.
  Rev. B}\ }\textbf {\bibinfo {volume} {69}},\ \bibinfo {pages} {113311}
  (\bibinfo {year} {2004})}\BibitemShut {NoStop}%
\bibitem [{\citenamefont {Nichele}\ \emph
  {et~al.}(2014{\natexlab{a}})\citenamefont {Nichele}, \citenamefont {Pal},
  \citenamefont {Winkler}, \citenamefont {Gerl}, \citenamefont {Wegscheider},
  \citenamefont {Ihn},\ and\ \citenamefont {Ensslin}}]{Nichele2014b}%
  \BibitemOpen
  \bibfield  {author} {\bibinfo {author} {\bibfnamefont {F.}~\bibnamefont
  {Nichele}}, \bibinfo {author} {\bibfnamefont {A.~N.}\ \bibnamefont {Pal}},
  \bibinfo {author} {\bibfnamefont {R.}~\bibnamefont {Winkler}}, \bibinfo
  {author} {\bibfnamefont {C.}~\bibnamefont {Gerl}}, \bibinfo {author}
  {\bibfnamefont {W.}~\bibnamefont {Wegscheider}}, \bibinfo {author}
  {\bibfnamefont {T.}~\bibnamefont {Ihn}}, \ and\ \bibinfo {author}
  {\bibfnamefont {K.}~\bibnamefont {Ensslin}},\ }\href
  {http://link.aps.org/doi/10.1103/PhysRevB.89.081306} {\bibfield  {journal}
  {\bibinfo  {journal} {Phys. Rev. B}\ }\textbf {\bibinfo {volume} {89}},\
  \bibinfo {pages} {081306} (\bibinfo {year} {2014}{\natexlab{a}})}\BibitemShut
  {NoStop}%
\bibitem [{\citenamefont {Minkov}\ \emph {et~al.}(2014)\citenamefont {Minkov},
  \citenamefont {Germanenko}, \citenamefont {Rut}, \citenamefont
  {Sherstobitov}, \citenamefont {Dvoretski},\ and\ \citenamefont
  {Mikhailov}}]{Minkov2014}%
  \BibitemOpen
  \bibfield  {author} {\bibinfo {author} {\bibfnamefont {G.~M.}\ \bibnamefont
  {Minkov}}, \bibinfo {author} {\bibfnamefont {A.~V.}\ \bibnamefont
  {Germanenko}}, \bibinfo {author} {\bibfnamefont {O.~E.}\ \bibnamefont {Rut}},
  \bibinfo {author} {\bibfnamefont {A.~A.}\ \bibnamefont {Sherstobitov}},
  \bibinfo {author} {\bibfnamefont {S.~A.}\ \bibnamefont {Dvoretski}}, \ and\
  \bibinfo {author} {\bibfnamefont {N.~N.}\ \bibnamefont {Mikhailov}},\ }\href
  {\doibase 10.1103/PhysRevB.89.165311} {\bibfield  {journal} {\bibinfo
  {journal} {Phys. Rev. B}\ }\textbf {\bibinfo {volume} {89}},\ \bibinfo
  {pages} {165311} (\bibinfo {year} {2014})}\BibitemShut {NoStop}%
\bibitem [{Sup()}]{Supplement}%
  \BibitemOpen
  \href@noop {} {}\bibinfo {note} {See Supplemental Material at [URL] for
  material and methods and a description of the numerical simulations, which
  includes
  Refs.~\cite{Nguyen2015,Tukey1967,Harris1978,Luryi1988,Kane1982,Bastard1988,Burt1992,Foreman1997,Novik2005,Pfeuffer-Jeschke,Halvorsen2000,Lawaetz1971,Andlauer2009}.}\BibitemShut
  {Stop}%
\bibitem [{\citenamefont {Zakharova}\ \emph {et~al.}(2002)\citenamefont
  {Zakharova}, \citenamefont {Yen},\ and\ \citenamefont
  {Chao}}]{Zakharova2002}%
  \BibitemOpen
  \bibfield  {author} {\bibinfo {author} {\bibfnamefont {A.}~\bibnamefont
  {Zakharova}}, \bibinfo {author} {\bibfnamefont {S.~T.}\ \bibnamefont {Yen}},
  \ and\ \bibinfo {author} {\bibfnamefont {K.~A.}\ \bibnamefont {Chao}},\
  }\href {\doibase 10.1103/PhysRevB.66.085312} {\bibfield  {journal} {\bibinfo
  {journal} {Phys. Rev. B}\ }\textbf {\bibinfo {volume} {66}},\ \bibinfo
  {pages} {085312} (\bibinfo {year} {2002})}\BibitemShut {NoStop}%
\bibitem [{\citenamefont {Li}\ \emph {et~al.}(2009)\citenamefont {Li},
  \citenamefont {Yang},\ and\ \citenamefont {Chang}}]{Li2009}%
  \BibitemOpen
  \bibfield  {author} {\bibinfo {author} {\bibfnamefont {J.}~\bibnamefont
  {Li}}, \bibinfo {author} {\bibfnamefont {W.}~\bibnamefont {Yang}}, \ and\
  \bibinfo {author} {\bibfnamefont {K.}~\bibnamefont {Chang}},\ }\href
  {\doibase 10.1103/PhysRevB.80.035303} {\bibfield  {journal} {\bibinfo
  {journal} {Phys. Rev. B}\ }\textbf {\bibinfo {volume} {80}},\ \bibinfo
  {pages} {035303} (\bibinfo {year} {2009})}\BibitemShut {NoStop}%
\bibitem [{\citenamefont {{Hu}}\ \emph {et~al.}(2016)\citenamefont {{Hu}},
  \citenamefont {{Liu}}, \citenamefont {{Xu}}, \citenamefont {{Zhang}},\ and\
  \citenamefont {{Zhou}}}]{Hu2016}%
  \BibitemOpen
  \bibfield  {author} {\bibinfo {author} {\bibfnamefont {L.-H.}\ \bibnamefont
  {{Hu}}}, \bibinfo {author} {\bibfnamefont {C.-X.}\ \bibnamefont {{Liu}}},
  \bibinfo {author} {\bibfnamefont {D.-H.}\ \bibnamefont {{Xu}}}, \bibinfo
  {author} {\bibfnamefont {F.-C.}\ \bibnamefont {{Zhang}}}, \ and\ \bibinfo
  {author} {\bibfnamefont {Y.}~\bibnamefont {{Zhou}}},\ }\href@noop {}
  {\bibfield  {journal} {\bibinfo  {journal} {ArXiv e-prints}\ } (\bibinfo
  {year} {2016})},\ \Eprint {http://arxiv.org/abs/1603.06074} {arXiv:1603.06074
  [cond-mat.mes-hall]} \BibitemShut {NoStop}%
\bibitem [{\citenamefont {Knez}\ \emph {et~al.}(2010)\citenamefont {Knez},
  \citenamefont {Du},\ and\ \citenamefont {Sullivan}}]{Knez2010}%
  \BibitemOpen
  \bibfield  {author} {\bibinfo {author} {\bibfnamefont {I.}~\bibnamefont
  {Knez}}, \bibinfo {author} {\bibfnamefont {R.~R.}\ \bibnamefont {Du}}, \ and\
  \bibinfo {author} {\bibfnamefont {G.}~\bibnamefont {Sullivan}},\ }\href
  {http://link.aps.org/doi/10.1103/PhysRevB.81.201301} {\bibfield  {journal}
  {\bibinfo  {journal} {Phys. Rev. B}\ }\textbf {\bibinfo {volume} {81}},\
  \bibinfo {pages} {201301} (\bibinfo {year} {2010})}\BibitemShut {NoStop}%
\bibitem [{\citenamefont {Nicholas}\ \emph {et~al.}(2000)\citenamefont
  {Nicholas}, \citenamefont {Takashina}, \citenamefont {Lakrimi}, \citenamefont
  {Kardynal}, \citenamefont {Khym}, \citenamefont {Mason}, \citenamefont
  {Symons}, \citenamefont {Maude},\ and\ \citenamefont
  {Portal}}]{Nicholas2000}%
  \BibitemOpen
  \bibfield  {author} {\bibinfo {author} {\bibfnamefont {R.~J.}\ \bibnamefont
  {Nicholas}}, \bibinfo {author} {\bibfnamefont {K.}~\bibnamefont {Takashina}},
  \bibinfo {author} {\bibfnamefont {M.}~\bibnamefont {Lakrimi}}, \bibinfo
  {author} {\bibfnamefont {B.}~\bibnamefont {Kardynal}}, \bibinfo {author}
  {\bibfnamefont {S.}~\bibnamefont {Khym}}, \bibinfo {author} {\bibfnamefont
  {N.~J.}\ \bibnamefont {Mason}}, \bibinfo {author} {\bibfnamefont {D.~M.}\
  \bibnamefont {Symons}}, \bibinfo {author} {\bibfnamefont {D.~K.}\
  \bibnamefont {Maude}}, \ and\ \bibinfo {author} {\bibfnamefont {J.~C.}\
  \bibnamefont {Portal}},\ }\href
  {http://link.aps.org/doi/10.1103/PhysRevLett.85.2364} {\bibfield  {journal}
  {\bibinfo  {journal} {Phys. Rev. Lett.}\ }\textbf {\bibinfo {volume} {85}},\
  \bibinfo {pages} {2364} (\bibinfo {year} {2000})}\BibitemShut {NoStop}%
\bibitem [{\citenamefont {Nichele}\ \emph
  {et~al.}(2014{\natexlab{b}})\citenamefont {Nichele}, \citenamefont {Pal},
  \citenamefont {Pietsch}, \citenamefont {Ihn}, \citenamefont {Ensslin},
  \citenamefont {Charpentier},\ and\ \citenamefont
  {Wegscheider}}]{Nichele2014}%
  \BibitemOpen
  \bibfield  {author} {\bibinfo {author} {\bibfnamefont {F.}~\bibnamefont
  {Nichele}}, \bibinfo {author} {\bibfnamefont {A.~N.}\ \bibnamefont {Pal}},
  \bibinfo {author} {\bibfnamefont {P.}~\bibnamefont {Pietsch}}, \bibinfo
  {author} {\bibfnamefont {T.}~\bibnamefont {Ihn}}, \bibinfo {author}
  {\bibfnamefont {K.}~\bibnamefont {Ensslin}}, \bibinfo {author} {\bibfnamefont
  {C.}~\bibnamefont {Charpentier}}, \ and\ \bibinfo {author} {\bibfnamefont
  {W.}~\bibnamefont {Wegscheider}},\ }\href {\doibase
  10.1103/PhysRevLett.112.036802} {\bibfield  {journal} {\bibinfo  {journal}
  {Phys. Rev. Lett.}\ }\textbf {\bibinfo {volume} {112}},\ \bibinfo {pages}
  {036802} (\bibinfo {year} {2014}{\natexlab{b}})}\BibitemShut {NoStop}%
\bibitem [{\citenamefont {Tsay}\ \emph {et~al.}(1997)\citenamefont {Tsay},
  \citenamefont {Chiang}, \citenamefont {Chau},\ and\ \citenamefont
  {Lo}}]{Tsay1997}%
  \BibitemOpen
  \bibfield  {author} {\bibinfo {author} {\bibfnamefont {S.-F.}\ \bibnamefont
  {Tsay}}, \bibinfo {author} {\bibfnamefont {J.-C.}\ \bibnamefont {Chiang}},
  \bibinfo {author} {\bibfnamefont {Z.~M.}\ \bibnamefont {Chau}}, \ and\
  \bibinfo {author} {\bibfnamefont {I.}~\bibnamefont {Lo}},\ }\href
  {http://link.aps.org/doi/10.1103/PhysRevB.56.13242} {\bibfield  {journal}
  {\bibinfo  {journal} {Phys. Rev. B}\ }\textbf {\bibinfo {volume} {56}},\
  \bibinfo {pages} {13242} (\bibinfo {year} {1997})}\BibitemShut {NoStop}%
\bibitem [{\citenamefont {Vasilyev}\ \emph {et~al.}(1999)\citenamefont
  {Vasilyev}, \citenamefont {Suchalkin}, \citenamefont {von Klitzing},
  \citenamefont {Meltser}, \citenamefont {Ivanov},\ and\ \citenamefont
  {Kop’ev}}]{Vasilyev1999}%
  \BibitemOpen
  \bibfield  {author} {\bibinfo {author} {\bibfnamefont {Y.}~\bibnamefont
  {Vasilyev}}, \bibinfo {author} {\bibfnamefont {S.}~\bibnamefont {Suchalkin}},
  \bibinfo {author} {\bibfnamefont {K.}~\bibnamefont {von Klitzing}}, \bibinfo
  {author} {\bibfnamefont {B.}~\bibnamefont {Meltser}}, \bibinfo {author}
  {\bibfnamefont {S.}~\bibnamefont {Ivanov}}, \ and\ \bibinfo {author}
  {\bibfnamefont {P.}~\bibnamefont {Kop’ev}},\ }\href
  {http://link.aps.org/doi/10.1103/PhysRevB.60.10636} {\bibfield  {journal}
  {\bibinfo  {journal} {Phys. Rev. B}\ }\textbf {\bibinfo {volume} {60}},\
  \bibinfo {pages} {10636} (\bibinfo {year} {1999})}\BibitemShut {NoStop}%
\bibitem [{\citenamefont {Ando}\ \emph {et~al.}(1982)\citenamefont {Ando},
  \citenamefont {Fowler},\ and\ \citenamefont {Stern}}]{Ando1982}%
  \BibitemOpen
  \bibfield  {author} {\bibinfo {author} {\bibfnamefont {T.}~\bibnamefont
  {Ando}}, \bibinfo {author} {\bibfnamefont {A.~B.}\ \bibnamefont {Fowler}}, \
  and\ \bibinfo {author} {\bibfnamefont {F.}~\bibnamefont {Stern}},\ }\href
  {\doibase 10.1103/RevModPhys.54.437} {\bibfield  {journal} {\bibinfo
  {journal} {Rev. Mod. Phys.}\ }\textbf {\bibinfo {volume} {54}},\ \bibinfo
  {pages} {437} (\bibinfo {year} {1982})}\BibitemShut {NoStop}%
\bibitem [{\citenamefont {Coleridge}\ \emph {et~al.}(1989)\citenamefont
  {Coleridge}, \citenamefont {Stoner},\ and\ \citenamefont
  {Fletcher}}]{Coleridge1989}%
  \BibitemOpen
  \bibfield  {author} {\bibinfo {author} {\bibfnamefont {P.~T.}\ \bibnamefont
  {Coleridge}}, \bibinfo {author} {\bibfnamefont {R.}~\bibnamefont {Stoner}}, \
  and\ \bibinfo {author} {\bibfnamefont {R.}~\bibnamefont {Fletcher}},\ }\href
  {http://link.aps.org/doi/10.1103/PhysRevB.39.1120} {\bibfield  {journal}
  {\bibinfo  {journal} {Phys. Rev. B}\ }\textbf {\bibinfo {volume} {39}},\
  \bibinfo {pages} {1120} (\bibinfo {year} {1989})}\BibitemShut {NoStop}%
\bibitem [{\citenamefont {Bychkov}\ and\ \citenamefont
  {Rashba}(1984)}]{Bychkov1984}%
  \BibitemOpen
  \bibfield  {author} {\bibinfo {author} {\bibfnamefont {Y.~A.}\ \bibnamefont
  {Bychkov}}\ and\ \bibinfo {author} {\bibfnamefont {E.~I.}\ \bibnamefont
  {Rashba}},\ }\href {http://stacks.iop.org/0022-3719/17/i=33/a=015} {\bibfield
   {journal} {\bibinfo  {journal} {Journal of Physics C: Solid State Physics}\
  }\textbf {\bibinfo {volume} {17}},\ \bibinfo {pages} {6039} (\bibinfo {year}
  {1984})}\BibitemShut {NoStop}%
\bibitem [{\citenamefont {Huckestein}(1995)}]{Huckestein1995}%
  \BibitemOpen
  \bibfield  {author} {\bibinfo {author} {\bibfnamefont {B.}~\bibnamefont
  {Huckestein}},\ }\href {\doibase 10.1103/RevModPhys.67.357} {\bibfield
  {journal} {\bibinfo  {journal} {Rev. Mod. Phys.}\ }\textbf {\bibinfo {volume}
  {67}},\ \bibinfo {pages} {357} (\bibinfo {year} {1995})}\BibitemShut
  {NoStop}%
\bibitem [{\citenamefont {Novoselov}\ \emph {et~al.}(2005)\citenamefont
  {Novoselov}, \citenamefont {Geim}, \citenamefont {Morozov}, \citenamefont
  {Jiang}, \citenamefont {Katsnelson}, \citenamefont {Grigorieva},
  \citenamefont {Dubonos},\ and\ \citenamefont {Firsov}}]{Novoselov2005}%
  \BibitemOpen
  \bibfield  {author} {\bibinfo {author} {\bibfnamefont {K.~S.}\ \bibnamefont
  {Novoselov}}, \bibinfo {author} {\bibfnamefont {A.~K.}\ \bibnamefont {Geim}},
  \bibinfo {author} {\bibfnamefont {S.~V.}\ \bibnamefont {Morozov}}, \bibinfo
  {author} {\bibfnamefont {D.}~\bibnamefont {Jiang}}, \bibinfo {author}
  {\bibfnamefont {M.~I.}\ \bibnamefont {Katsnelson}}, \bibinfo {author}
  {\bibfnamefont {I.~V.}\ \bibnamefont {Grigorieva}}, \bibinfo {author}
  {\bibfnamefont {S.~V.}\ \bibnamefont {Dubonos}}, \ and\ \bibinfo {author}
  {\bibfnamefont {A.~A.}\ \bibnamefont {Firsov}},\ }\href
  {http://dx.doi.org/10.1038/nature04233} {\bibfield  {journal} {\bibinfo
  {journal} {Nature}\ }\textbf {\bibinfo {volume} {438}},\ \bibinfo {pages}
  {197} (\bibinfo {year} {2005})}\BibitemShut {NoStop}%
\bibitem [{\citenamefont {Zhang}\ \emph {et~al.}(2005)\citenamefont {Zhang},
  \citenamefont {Tan}, \citenamefont {Stormer},\ and\ \citenamefont
  {Kim}}]{Zhang2005}%
  \BibitemOpen
  \bibfield  {author} {\bibinfo {author} {\bibfnamefont {Y.}~\bibnamefont
  {Zhang}}, \bibinfo {author} {\bibfnamefont {Y.-W.}\ \bibnamefont {Tan}},
  \bibinfo {author} {\bibfnamefont {H.~L.}\ \bibnamefont {Stormer}}, \ and\
  \bibinfo {author} {\bibfnamefont {P.}~\bibnamefont {Kim}},\ }\href
  {http://dx.doi.org/10.1038/nature04235} {\bibfield  {journal} {\bibinfo
  {journal} {Nature}\ }\textbf {\bibinfo {volume} {438}},\ \bibinfo {pages}
  {201} (\bibinfo {year} {2005})}\BibitemShut {NoStop}%
\bibitem [{\citenamefont {Qu}\ \emph {et~al.}(2010)\citenamefont {Qu},
  \citenamefont {Hor}, \citenamefont {Xiong}, \citenamefont {Cava},\ and\
  \citenamefont {Ong}}]{Qu2010}%
  \BibitemOpen
  \bibfield  {author} {\bibinfo {author} {\bibfnamefont {D.-X.}\ \bibnamefont
  {Qu}}, \bibinfo {author} {\bibfnamefont {Y.~S.}\ \bibnamefont {Hor}},
  \bibinfo {author} {\bibfnamefont {J.}~\bibnamefont {Xiong}}, \bibinfo
  {author} {\bibfnamefont {R.~J.}\ \bibnamefont {Cava}}, \ and\ \bibinfo
  {author} {\bibfnamefont {N.~P.}\ \bibnamefont {Ong}},\ }\href {\doibase
  10.1126/science.1189792} {\bibfield  {journal} {\bibinfo  {journal}
  {Science}\ }\textbf {\bibinfo {volume} {329}},\ \bibinfo {pages} {821}
  (\bibinfo {year} {2010})}\BibitemShut {NoStop}%
\bibitem [{\citenamefont {Xiong}\ \emph {et~al.}(2012)\citenamefont {Xiong},
  \citenamefont {Luo}, \citenamefont {Khoo}, \citenamefont {Jia}, \citenamefont
  {Cava},\ and\ \citenamefont {Ong}}]{Xiong2012}%
  \BibitemOpen
  \bibfield  {author} {\bibinfo {author} {\bibfnamefont {J.}~\bibnamefont
  {Xiong}}, \bibinfo {author} {\bibfnamefont {Y.}~\bibnamefont {Luo}}, \bibinfo
  {author} {\bibfnamefont {Y.}~\bibnamefont {Khoo}}, \bibinfo {author}
  {\bibfnamefont {S.}~\bibnamefont {Jia}}, \bibinfo {author} {\bibfnamefont
  {R.~J.}\ \bibnamefont {Cava}}, \ and\ \bibinfo {author} {\bibfnamefont
  {N.~P.}\ \bibnamefont {Ong}},\ }\href {\doibase 10.1103/PhysRevB.86.045314}
  {\bibfield  {journal} {\bibinfo  {journal} {Phys. Rev. B}\ }\textbf {\bibinfo
  {volume} {86}},\ \bibinfo {pages} {045314} (\bibinfo {year}
  {2012})}\BibitemShut {NoStop}%
\bibitem [{\citenamefont {Wright}\ and\ \citenamefont
  {McKenzie}(2013)}]{Wright2013}%
  \BibitemOpen
  \bibfield  {author} {\bibinfo {author} {\bibfnamefont {A.~R.}\ \bibnamefont
  {Wright}}\ and\ \bibinfo {author} {\bibfnamefont {R.~H.}\ \bibnamefont
  {McKenzie}},\ }\href {\doibase 10.1103/PhysRevB.87.085411} {\bibfield
  {journal} {\bibinfo  {journal} {Phys. Rev. B}\ }\textbf {\bibinfo {volume}
  {87}},\ \bibinfo {pages} {085411} (\bibinfo {year} {2013})}\BibitemShut
  {NoStop}%
\bibitem [{\citenamefont {Nguyen}\ \emph {et~al.}(2015)\citenamefont {Nguyen},
  \citenamefont {Yi}, \citenamefont {Noah}, \citenamefont {Thorp},\ and\
  \citenamefont {Sokolich}}]{Nguyen2015}%
  \BibitemOpen
  \bibfield  {author} {\bibinfo {author} {\bibfnamefont {B.-M.}\ \bibnamefont
  {Nguyen}}, \bibinfo {author} {\bibfnamefont {W.}~\bibnamefont {Yi}}, \bibinfo
  {author} {\bibfnamefont {R.}~\bibnamefont {Noah}}, \bibinfo {author}
  {\bibfnamefont {J.}~\bibnamefont {Thorp}}, \ and\ \bibinfo {author}
  {\bibfnamefont {M.}~\bibnamefont {Sokolich}},\ }\href {\doibase
  http://dx.doi.org/10.1063/1.4906589} {\bibfield  {journal} {\bibinfo
  {journal} {Applied Physics Letters}\ }\textbf {\bibinfo {volume} {106}},\
  \bibinfo {pages} {032107} (\bibinfo {year} {2015})}\BibitemShut {NoStop}%
\bibitem [{\citenamefont {Tukey}(1967)}]{Tukey1967}%
  \BibitemOpen
  \bibfield  {author} {\bibinfo {author} {\bibfnamefont {J.}~\bibnamefont
  {Tukey}},\ }\href {http://books.google.it/books?id=7Mg-AAAAIAAJ} {\emph
  {\bibinfo {title} {Advance Seminar on Spectral analysis of time series:
  proceedings}}}\ (\bibinfo  {publisher} {Wiley},\ \bibinfo {year} {1967})\
  pp.\ \bibinfo {pages} {25--46}\BibitemShut {NoStop}%
\bibitem [{\citenamefont {Harris}(1978)}]{Harris1978}%
  \BibitemOpen
  \bibfield  {author} {\bibinfo {author} {\bibfnamefont {F.}~\bibnamefont
  {Harris}},\ }\bibfield  {booktitle} {\emph {\bibinfo {booktitle} {Proceedings
  of the IEEE}},\ }\href@noop {} {\bibfield  {journal} {\bibinfo  {journal}
  {Proc. IEEE}\ }\textbf {\bibinfo {volume} {66}},\ \bibinfo {pages} {51}
  (\bibinfo {year} {1978})}\BibitemShut {NoStop}%
\bibitem [{\citenamefont {Luryi}(1988)}]{Luryi1988}%
  \BibitemOpen
  \bibfield  {author} {\bibinfo {author} {\bibfnamefont {S.}~\bibnamefont
  {Luryi}},\ }\href {\doibase http://dx.doi.org/10.1063/1.99649} {\bibfield
  {journal} {\bibinfo  {journal} {Appl. Phys. Lett.}\ }\textbf {\bibinfo
  {volume} {52}},\ \bibinfo {pages} {501} (\bibinfo {year}
  {Luryi1988})}\BibitemShut {NoStop}%
\bibitem [{\citenamefont {Kane}(1982)}]{Kane1982}%
  \BibitemOpen
  \bibfield  {author} {\bibinfo {author} {\bibfnamefont {E.~O.}\ \bibnamefont
  {Kane}},\ }in\ \href@noop {} {\emph {\bibinfo {booktitle} {Handbook on
  Semiconductors}}},\ Vol.~\bibinfo {volume} {1},\ \bibinfo {editor} {edited
  by\ \bibinfo {editor} {\bibfnamefont {W.}~\bibnamefont {Paul}}}\ (\bibinfo
  {publisher} {North-Holland, Amsterdam},\ \bibinfo {year} {1982})\ p.\
  \bibinfo {pages} {193}\BibitemShut {NoStop}%
\bibitem [{\citenamefont {Bastard}(1988)}]{Bastard1988}%
  \BibitemOpen
  \bibfield  {author} {\bibinfo {author} {\bibfnamefont {G.}~\bibnamefont
  {Bastard}},\ }\href@noop {} {\emph {\bibinfo {title} {Wave Mechanics Applied
  to Semiconductor Heterostructures}}}\ (\bibinfo  {publisher} {Wiley, New
  York},\ \bibinfo {year} {1988})\BibitemShut {NoStop}%
\bibitem [{\citenamefont {Burt}(1992)}]{Burt1992}%
  \BibitemOpen
  \bibfield  {author} {\bibinfo {author} {\bibfnamefont {M.~G.}\ \bibnamefont
  {Burt}},\ }\href {http://stacks.iop.org/0953-8984/4/i=32/a=003} {\bibfield
  {journal} {\bibinfo  {journal} {J. Phys.: Condens. Matter}\ }\textbf
  {\bibinfo {volume} {4}},\ \bibinfo {pages} {6651} (\bibinfo {year}
  {1992})}\BibitemShut {NoStop}%
\bibitem [{\citenamefont {Foreman}(1997)}]{Foreman1997}%
  \BibitemOpen
  \bibfield  {author} {\bibinfo {author} {\bibfnamefont {B.~A.}\ \bibnamefont
  {Foreman}},\ }\href {\doibase 10.1103/PhysRevB.56.R12748} {\bibfield
  {journal} {\bibinfo  {journal} {Phys. Rev. B}\ }\textbf {\bibinfo {volume}
  {56}},\ \bibinfo {pages} {R12748} (\bibinfo {year} {1997})}\BibitemShut
  {NoStop}%
\bibitem [{\citenamefont {Novik}\ \emph {et~al.}(2005)\citenamefont {Novik},
  \citenamefont {Pfeuffer-Jeschke}, \citenamefont {Jungwirth}, \citenamefont
  {Latussek}, \citenamefont {Becker}, \citenamefont {Landwehr}, \citenamefont
  {Buhmann},\ and\ \citenamefont {Molenkamp}}]{Novik2005}%
  \BibitemOpen
  \bibfield  {author} {\bibinfo {author} {\bibfnamefont {E.~G.}\ \bibnamefont
  {Novik}}, \bibinfo {author} {\bibfnamefont {A.}~\bibnamefont
  {Pfeuffer-Jeschke}}, \bibinfo {author} {\bibfnamefont {T.}~\bibnamefont
  {Jungwirth}}, \bibinfo {author} {\bibfnamefont {V.}~\bibnamefont {Latussek}},
  \bibinfo {author} {\bibfnamefont {C.~R.}\ \bibnamefont {Becker}}, \bibinfo
  {author} {\bibfnamefont {G.}~\bibnamefont {Landwehr}}, \bibinfo {author}
  {\bibfnamefont {H.}~\bibnamefont {Buhmann}}, \ and\ \bibinfo {author}
  {\bibfnamefont {L.~W.}\ \bibnamefont {Molenkamp}},\ }\href {\doibase
  10.1103/PhysRevB.72.035321} {\bibfield  {journal} {\bibinfo  {journal} {Phys.
  Rev. B}\ }\textbf {\bibinfo {volume} {72}},\ \bibinfo {pages} {035321}
  (\bibinfo {year} {2005})}\BibitemShut {NoStop}%
\bibitem [{\citenamefont {Pfeuffer-Jeschke}(2000)}]{Pfeuffer-Jeschke}%
  \BibitemOpen
  \bibfield  {author} {\bibinfo {author} {\bibfnamefont {A.~J.}\ \bibnamefont
  {Pfeuffer-Jeschke}},\ }\emph {\bibinfo {title} {Transport experiments in
  two-dimensional systems with strong spin-orbit interaction}},\ \href@noop {}
  {Ph.D. thesis},\ \bibinfo  {school} {Physikalisches Institut, Universit\"at
  W\"urzburg} (\bibinfo {year} {2000})\BibitemShut {NoStop}%
\bibitem [{\citenamefont {Halvorsen}\ \emph {et~al.}(2000)\citenamefont
  {Halvorsen}, \citenamefont {Galperin},\ and\ \citenamefont
  {Chao}}]{Halvorsen2000}%
  \BibitemOpen
  \bibfield  {author} {\bibinfo {author} {\bibfnamefont {E.}~\bibnamefont
  {Halvorsen}}, \bibinfo {author} {\bibfnamefont {Y.}~\bibnamefont {Galperin}},
  \ and\ \bibinfo {author} {\bibfnamefont {K.~A.}\ \bibnamefont {Chao}},\
  }\href {\doibase 10.1103/PhysRevB.61.16743} {\bibfield  {journal} {\bibinfo
  {journal} {Phys. Rev. B}\ }\textbf {\bibinfo {volume} {61}},\ \bibinfo
  {pages} {16743} (\bibinfo {year} {2000})}\BibitemShut {NoStop}%
\bibitem [{\citenamefont {Lawaetz}(1971)}]{Lawaetz1971}%
  \BibitemOpen
  \bibfield  {author} {\bibinfo {author} {\bibfnamefont {P.}~\bibnamefont
  {Lawaetz}},\ }\href {\doibase 10.1103/PhysRevB.4.3460} {\bibfield  {journal}
  {\bibinfo  {journal} {Phys. Rev. B}\ }\textbf {\bibinfo {volume} {4}},\
  \bibinfo {pages} {3460} (\bibinfo {year} {1971})}\BibitemShut {NoStop}%
\bibitem [{\citenamefont {Andlauer}\ and\ \citenamefont
  {Vogl}(2009)}]{Andlauer2009}%
  \BibitemOpen
  \bibfield  {author} {\bibinfo {author} {\bibfnamefont {T.}~\bibnamefont
  {Andlauer}}\ and\ \bibinfo {author} {\bibfnamefont {P.}~\bibnamefont
  {Vogl}},\ }\href {\doibase 10.1103/PhysRevB.80.035304} {\bibfield  {journal}
  {\bibinfo  {journal} {Phys. Rev. B}\ }\textbf {\bibinfo {volume} {80}},\
  \bibinfo {pages} {035304} (\bibinfo {year} {2009})}\BibitemShut {NoStop}%
\end{thebibliography}%
\newpage

\section{Supplemental Material}
This Supplemental Material Section describes the wafer structure, the sample fabrication procedure and the measuring setup. We describe the numerical procedure used to Fourier transform the magnetoresistance and derive the subbands density. We further describe the electrostatic model used to calculate electron and hole densities as a function of top gate voltage and the $\boldsymbol{k}\cdot\boldsymbol{p}$ simulations for band structure calculations.




\setcounter{equation}{0}
\renewcommand{\theequation}{S.\arabic{equation}}
\setcounter{figure}{0}
\renewcommand{\thefigure}{S.\arabic{figure}}

\subsection{Material and Methods}
The wafer structure was grown by molecular beam epitaxy on a [001] oriented GaSb substrate. From top to bottom it consists of a $3~\rm{nm}$ GaSb capping layer, a $50~\rm{nm}$ AlSb insulating barrier, a $5~\rm{nm}$ GaSb QW grown on top of a $12.5~\rm{nm}$ InAs QW, a second AlSb barrier and a thick GaSb buffer layer. More information on wafer growth are reported in Ref.~\onlinecite{Nguyen2015,Qu2015,Nichele2015}.

A $100\times 50 ~\rm{\mu m^2}$ Hall bar structure was patterned with conventional electron beam lithography techniques and wet etching. The Hall bar structure was oriented along the [110] crystallographic direction. For wet etching we used a general III-V etching solution consisting of $\rm{H_2O:C_6H_8O_7:H_3PO_4:H_2O_2}$ in concentration $220:55:3:3$. The solution was kept at room temperature and well stirred, resulting in an etching rate of approximately $1~\rm{nm~s^{-1}}$. Ohmic contacts were defined by etching the wafer down to the InAs quantum well and depositing Ti/Au electrodes, without any annealing. The sample was covered with a $40~\rm{nm}$ $\rm{HfO_2}$ insulating layer grown by atomic layer deposition and a global Ti/Au top gate.

Transport measurements were performed in a dilution refrigerator with a base temperature of $50~\rm{mK}$ using low frequency ($<100~\rm{Hz}$) lock-in techniques. The amplitude of the AC currents was always kept small enough ($\leq 20~\rm{nA}$) to prevent sample heating. Due to the onset of leakage currents at finite bias, the device was operated at zero back gate voltage, where the resistance between the 2DEG and the back gate was in excess of $10~\rm{G\Omega}$.

\subsection{Fourier transforming techniques}
We now describe the numerical procedure used to Fourier transform the Shubnikov-de Haas (SdH) oscillations in the longitudinal resistivity $\rho_{xx}$. The magnetic field range for the analysis was chosen, case by case, to include only SdH oscillations whose amplitude was small compared to the zero field $\rho_{xx}$. We first plotted the quantity $\rho_{xx}(1/B)$, where the oscillations are periodic. The curve was interpolated on a new $1/B$ axis with constant spacing between points. At this point, we removed the slowly varying background of the data by subtracting the fit to a low order polynomial. In order to improve the final output of the Fourier transform, we adopted standard numerical procedures \cite{Tukey1967} such as padding the data with zeros and windowing \cite{Harris1978}. In case of a 2DEG with strong spin-orbit coupling (SOC), the frequency axis $f$ is converted into density via $n_{i}=ef_{i}/h$. In a conventional two-dimensional electron gas, an additional factor of two would be necessary to convert frequency into densities: $n_{i}=2ef_{i}/h$. The same analysis is described with greater details in Ref.~\onlinecite{Nichele2014b}. We note that the position of the power spectrum peaks of Fig.~3 of the main text was affected by less than $5\%$ by modifying the details of this analysis.  The final result of this procedure still contains spurious low frequency components, which originate from the difficulties in completely removing the slowly varying background of the original data. To suppress such features we multiplied the spectra by a high pass filter with cut-off frequency $ef/h=0.4\times 10^{15}$.
Finally, the spectra were normalized, for each value of $V_{\mathrm{TG}}$, to the amplitude of the $n_1$ peak.

The Fourier transform analysis presented in the main text for $V_{\mathrm{TG}}\geq-0.46V$ is fully compatible with two spin-orbit split subbands. In fact the relation $n_1+n_2=n_{\mathrm{Hall}}$ is always satisfied, where $n_{\mathrm{Hall}}$ is independently obtained from $\rho_{xy}$. In case of orbital electron-like subbands, an additional factor of two would be necessary for converting the power spectrum frequencies to densities, to include spin degeneracy.

\subsection{Additional Data}

\subsubsection{Charge Neutrality Point}
\begin{figure}
\includegraphics[width=\linewidth]{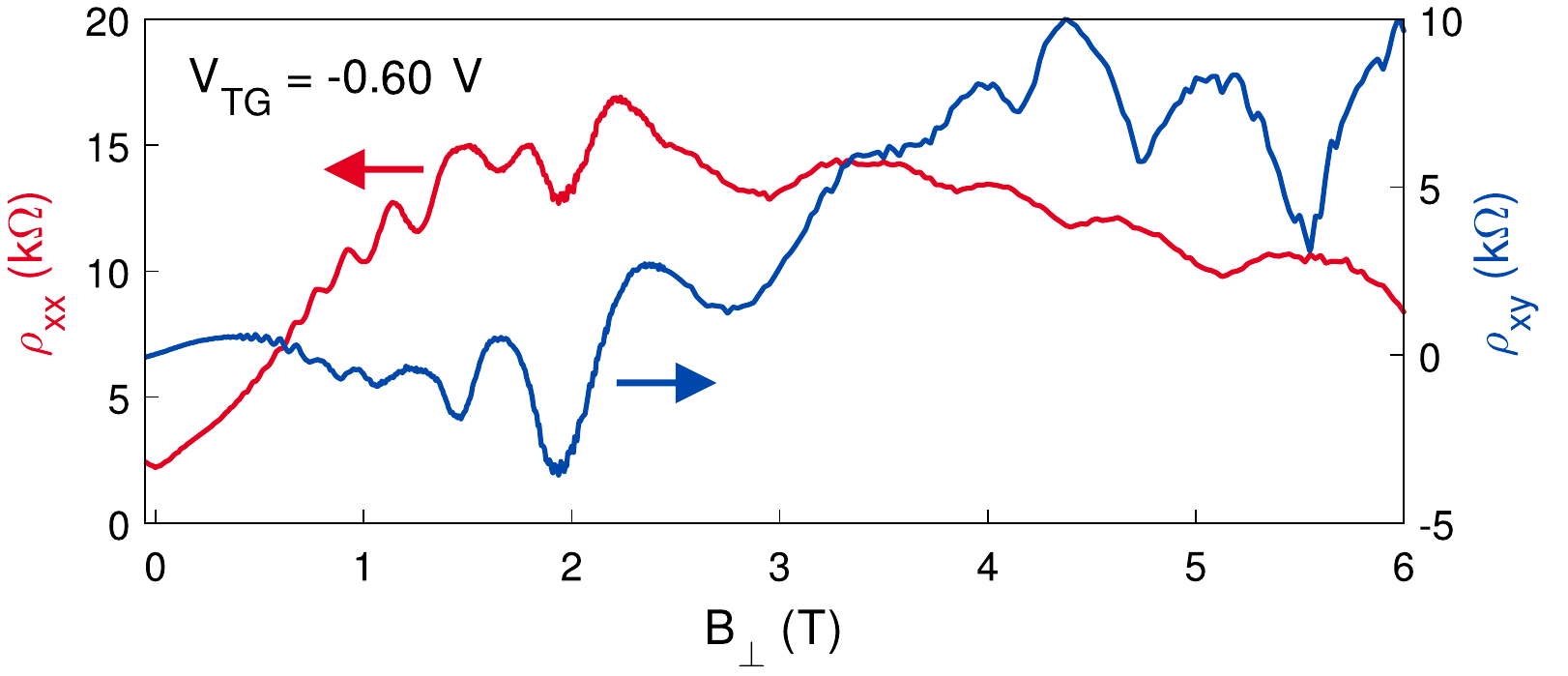}
\caption{Longitudinal (red) and transverse (blue) resistivities measured at the zero-field CNP.}
\label{fig:rxx}
\end{figure}

Figure~\ref{fig:rxx} shows in more detail the longitudinal resistivity $\rho_{xx}$ and the transverse resistivity $\rho_{xy}$ (red and blue respectively) measured at $V_{\mathrm{TG}}=-0.59~\mathrm{V}$, identified as the zero field CNP. The lack of a net slope in $\rho_{xy}$ is indicative of equal electron and hole concentration. The presence of oscillations in $\rho_{xx}$ confirms the dominant contribution of bulk transport at the CNP, as predicted by band structure calculations for negative gate voltages [cfg. Fig.~\ref{disps}(a)].

\subsubsection{Spin-Orbit Splitting in the Electron Regime}
\begin{figure}
\includegraphics[width=\linewidth]{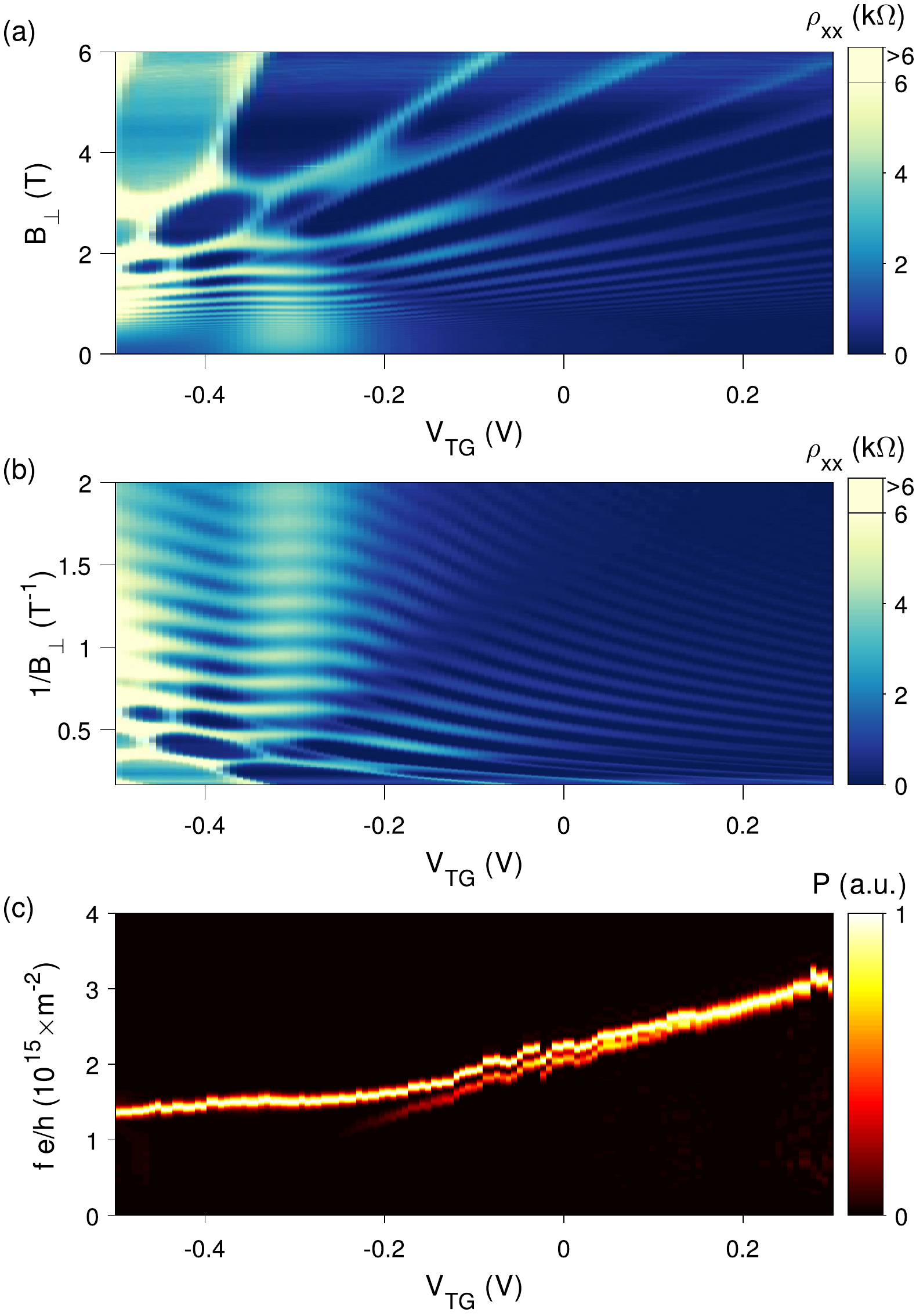}
\caption{Same data presented in the main text, but focused on the electron regime. (a) Longitudinal resistivity $\rho_{xx}$. (b) As in (a) but as a function of inverse out-of-plane magnetic field $B_{\perp}$. (c) Power spectrum of the data in (b). The amplitude has been normalized, column by column, to the $n_1$ peak. Differently from Fig.~3(c) of the main text, we plot the amplitude in linear scale.}
\label{fig:FFT}
\end{figure}

For better clarity, we reproduce here the longitudinal resistivity data measured in the electron regime without any overlay. Figure~\ref{fig:FFT}(a) shows the longitudinal resistivity $\rho_{xx}$ in the gate range where no indication of hole transport is detected. The same data is shown again in Fig.~\ref{fig:FFT}(b) plotted as a function of $1/B_{\perp}$, that makes the oscillations quasi-periodic along the vertical axis. Finally, Fig.~\ref{fig:FFT}(c) shows the power spectrum of the data in Fig.~\ref{fig:FFT}(b) in arbitrary units and linear scale. From the power spectrum we observe a splitting of the main peak, coincidental with the development of a beating pattern in $\rho_{xx}(1/B_{\perp})$. As the lower density split peak moves to lower density, its amplitude decreases.

\subsection{Numerical simulations}

\subsubsection{Capacitor model}

\begin{figure}[tb!]
\includegraphics[width=\linewidth]{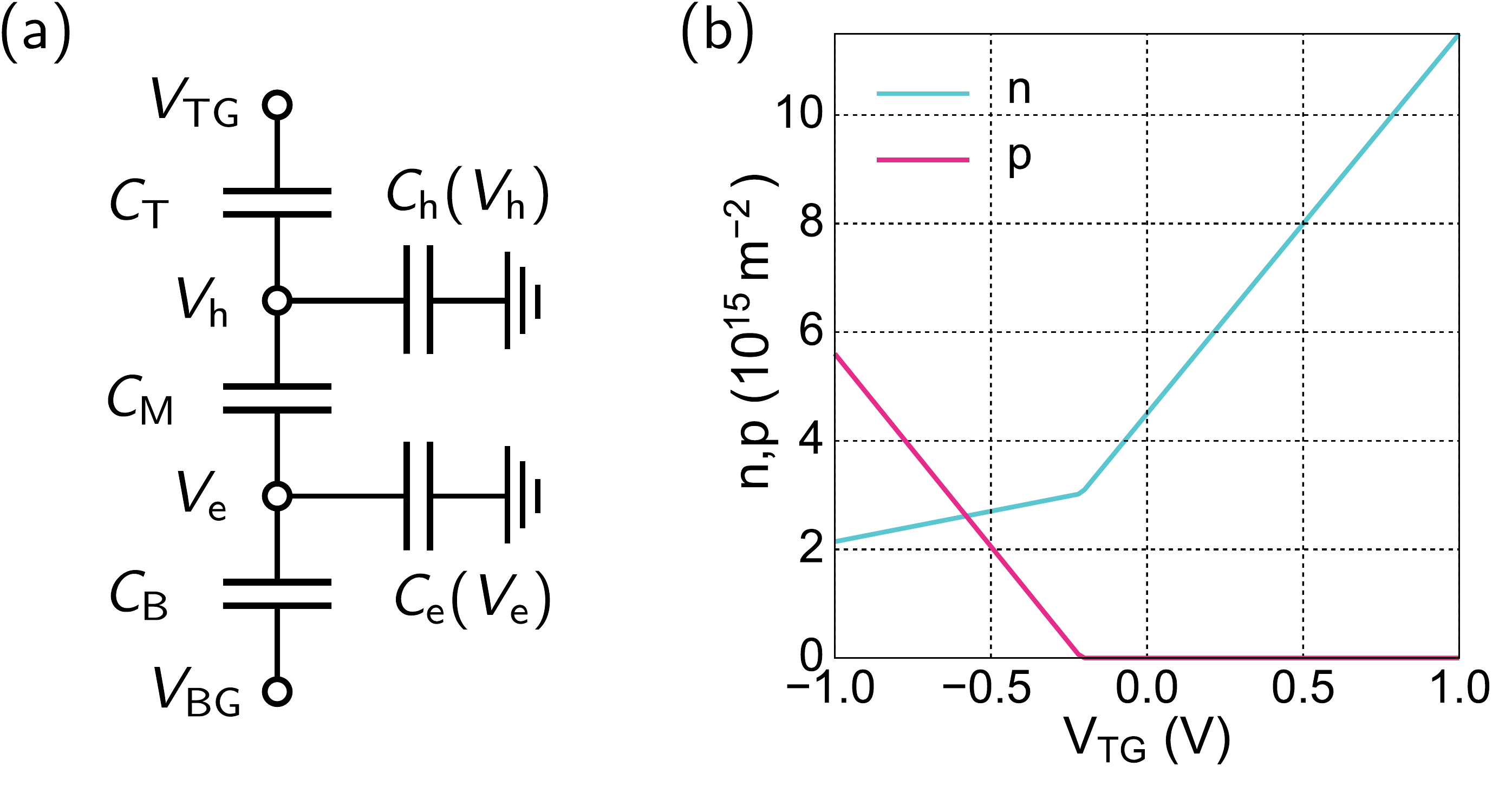}
\caption{(a) Schematic of the capacitor model for InAs/GaSb quantum wells. (b) Electron ($n$) and hole ($p$) densities obtained from the capacitor model as a function of the top-gate voltage $V_\text{TG}$ }
\label{fig:capmod}
\end{figure}

To estimate the electric fields in the quantum well we use the capacitor model introduced in Ref.~\onlinecite{Qu2015} (see Fig.~\ref{fig:capmod}(a). With the material parameters from Ref.~\onlinecite{Qu2015} and assuming the dielectric constant of HfO$_2$ as $\epsilon_\text{HfO$_2$} = 25$ we find the geometric capacitances for our quantum well structure as $C_\text{T} = 135$\,nF/cm$^2$, $C_\text{M} = 1.6\, \mu$F/cm$^2$, and $C_\text{B} = 177$\,nF/cm$^2$. The quantities $C_\text{e}$ and $C_\text{h}$ are quantum capacitances \cite{Luryi1988} that are non-zero only if there is a finite carrier density. The capacitor model neglects the intrinsic inversion of electron and hole bands in the InAs/GaSb quantum well, but assumes that the Fermi energy is aligned with the electron and hole band bottom when the potential in the respective layer is $0\,$V. Hence, $C_\text{e}= 2.7\,\mu$F/cm$^2$ if $V_e>0$ and zero else, whereas $C_\text{e}= 6\,\mu$F/cm$^2$ if $V_h<0$ and zero else. $V_\text{e}$ and $V_\text{h}$ are the potential values in the middle of the InAs and GaSb layers.

In our experiments, the back gate is always kept at $V_\text{BG} =0\,$V. The assumptions of the capacitor model then imply that for $V_\text{TG}=0\,$V both electron and hole density are zero.
In the experiment, we however find a nonzero electron density in this case, due to in-built electric fields. We approximate these electric fields by a finite fictitious back gate voltage $V_\text{BG}$ that we fix such that we recover the experimentally measured electron density of $n \approx 4.5 \times 10^{15} m^{-2}$ for $V_\text{TG} = 0\,$V (see Fig.~3 in the main text).

Even with its simplifications, the capacitor model captures essential features of the experiment: Fig.~\ref{fig:capmod}(b) shows electron and hole densities as a function of top gate voltage. Equal densities of electrons and holes are found around $V_\text{TG}=-0.6\,$V which agrees with the voltage where the charge neutrality point is found experimentally. Additionally, we observe that the gate-tunability of the electron density is strongly reduced when holes are occupied. The finite density of charge carriers in the GaSb (hole) layer lieing between top gate and (electron) InAs layer screens the electric field. A similar feature is seen in Fig.~3 of the main text. The screening by the hole layer also explains why experimentally the second electron Fermi surface is never recovered within our gate voltage range.

We note that the capacitor model assumes constant density of states of the electron and hole layer, and no electron-hole mixing. For this reason, the quantitative results of Fig.~\ref{fig:capmod} are valid in the high density limit, but should be taken with care close to the CNP, where the density of states shows gaps and singularities.

\subsubsection[$\boldsymbol{k}\cdot\boldsymbol{p}$ simulations]{k.p simulations}

The numerical band structure simulations use the standard semiconductor $\boldsymbol{k}\cdot\boldsymbol{p}$ model~\cite{Kane1982,Bastard1988}. The $8\times 8$ Kane Hamiltonian contains position-dependent
parameters corresponding to the different material layers, and must be properly symmetrized. Following the symmetrization put forward by Burt and Foreman\cite{Burt1992, Foreman1997}, the Hamiltonian for the [001] growth direction takes the following form:\cite{Novik2005, Pfeuffer-Jeschke}

\begingroup
\renewcommand*{\arraystretch}{2.5}
\begin{widetext}
\begin{equation}
\label{H8x8}
H=
\begin{pmatrix}
T & 0 & -\frac{1}{\sqrt{2}} P k_{+} & \sqrt{\frac{2}{3}} P k_{z} & \frac{1}{\sqrt{6}}P k_{-} & 0 & -\frac{1}{\sqrt{3}} P k_{z} & -\frac{1}{\sqrt{3}} P k_{-} \\
0 & T & 0 & -\frac{1}{\sqrt{6}}P k_{+} & \sqrt{\frac{2}{3}} P k_{z} & \frac{1}{\sqrt{2}} P k_{-} & -\frac{1}{\sqrt{3}} P k_{+} & \frac{1}{\sqrt{3}} P k_{z} \\
-\frac{1}{\sqrt{2}} k_{-} P & 0 & U+V & -\overline{S}_{-} & R & 0 & \frac{1}{\sqrt{2}}\overline{S}_{-} & -\sqrt{2} R \\
\sqrt{\frac{2}{3}}k_z P & -\frac{1}{\sqrt{6}} k_{-} P & -\overline{S}^{\dagger}_{-} & U-V & C & R & \sqrt{2} V & -\sqrt{\frac{3}{2}} \widetilde{S}_{-} \\
\frac{1}{\sqrt{6}}k_{+} P & \sqrt{\frac{2}{3}} k_z P & R^{\dagger} & C^{\dagger} & U-V & \overline{S}^{\dagger}_{+} & -\sqrt{\frac{3}{2}}\widetilde{S}_{+} & -\sqrt{2} V \\
0 & \frac{1}{\sqrt{2}} k_{+}P & 0 & R^{\dagger} & \overline{S}_{+} & U+V & \sqrt{2}R^{\dagger} & \frac{1}{\sqrt{2}}\overline{S}_{+} \\
-\frac{1}{\sqrt{3}} k_{z} P & -\frac{1}{\sqrt{3}} k_{-} P & \frac{1}{\sqrt{2}} \overline{S}^{\dagger}_{-} & \sqrt{2} V & -\sqrt{\frac{3}{2}} \widetilde{S}^{\dagger}_{+} & \sqrt{2} R & U - \Delta & C \\
-\frac{1}{\sqrt{3}} k_{+} P & \frac{1}{\sqrt{3}} k_{z} P & -\sqrt{2} R^{\dagger} & -\sqrt{\frac{3}{2}} \widetilde{S}^{\dagger}_{-} & -\sqrt{2} V & \frac{1}{\sqrt{2}} \overline{S}^{\dagger}_{+} & C^{\dagger} & U-\Delta
\end{pmatrix},
\end{equation}
where
\end{widetext}
\begin{equation*}
k_\parallel^2 = k_x^2 + k_y^2, \quad k_\pm = k_x \pm i k_y, \quad k_z = -i \partial / \partial z,
\end{equation*}
\begin{align*}
T &= E_c + \frac{\hbar^2}{2m_0} \left(\gamma^{\prime}_0k_\parallel^2 + k_z \gamma^{\prime}_0 k_z\right), \\
U &= E_v -\frac{\hbar^2}{2m_0} \left(\gamma^{\prime}_1 k_\parallel^2 + k_z \gamma^{\prime}_1 k_z \right), \\
V &= -\frac{\hbar^2}{2m_0} \left( \gamma^{\prime}_2 k_\parallel^2 - 2 k_z \gamma^{\prime}_2 k_z \right), \\
R &= -\frac{\hbar^2}{2m_0} \frac{\sqrt{3}}{2} \left[(\gamma^{\prime}_3-\gamma^{\prime}_2)k_{+}^2 - (\gamma^{\prime}_3+\gamma^{\prime}_2)k_{-}^2\right] ,\\
\overline{S}_{\pm} &= -\frac{\hbar^2}{2m_0} \sqrt{3} k_\pm \left(\{\gamma^{\prime}_3, k_z\} + [\kappa^{\prime}, k_z]\right),\\
\widetilde{S}_{\pm} &= -\frac{\hbar^2}{2m_0} \sqrt{3} k_\pm \left(\{\gamma^{\prime}_3, k_z\} - \frac{1}{3} [\kappa^{\prime}, k_z]\right), \\
C &= \frac{\hbar^2}{m_0}k_{-} \left[\kappa^{\prime}, k_z\right].
\end{align*}
\endgroup
Here, $P$ is the Kane momentum matrix element, $E_c$ and $E_v$ are the conduction and valence band edges, respectively, and $\Delta$ is the spin-orbit splitting energy. $[A, B] = AB - BA$ is the commutator and $\{A, B\} = AB + BA$ is the anticommutator for the operators A and B.

$\gamma^{\prime}_0$, $\gamma^{\prime}_1$, $\gamma^{\prime}_2$, $\gamma^{\prime}_3$ and $\kappa^{\prime}$ are the renormalized band parameters entering the $8\times 8$ Hamiltonian. They are related
to the effective mass of the conduction band ($m_c$) and the Luttinger parameters of the hole bands ($\gamma_{1,2,3}$ and $\kappa$) through
\begin{align}
\gamma^\prime_0 &= \gamma_0 - \frac{E_P}{E_g} \frac{E_g + \frac{2}{3}\Delta}{E_g + \Delta},\label{g0r}\\
\gamma^{\prime}_1 &= \gamma_1 - \frac{1}{3}\frac{E_P}{E_g},\label{g1r}\\
\gamma^{\prime}_2 &= \gamma_2 - \frac{1}{6}\frac{E_P}{E_g},\label{g2r}\\
\gamma^{\prime}_3 &= \gamma_3 - \frac{1}{6}\frac{E_P}{E_g}\label{g3r},\\
\kappa^{\prime} &= \kappa - \frac{1}{6}\frac{E_P}{E_g}\label{kr},
\end{align}
where
\begin{equation}
E_P = \frac{2m_0 P^2}{\hbar^2}, \quad \gamma_0 = \frac{m_0}{m_c},
\end{equation}
and $E_g$ is a band gap.

All of these parameters are material dependent and hence a function of the $z$-coordinate. The order of operators in \eqref{H8x8} is such that the Hamiltonian is indeed Hermitian.

The Hamiltonian~\eqref{H8x8} exhibits unphysical solutions inside the band gap if $\gamma_0' < 0$. The spurious solutions appear at large $\boldsymbol{k}$-values, beyond the validity of the $\boldsymbol{k}\cdot\boldsymbol{p}$-model. In order to avoid these unphysical states, we apply the method put forward in Ref.~\onlinecite{Foreman1997}: we renormalize $P$ in a way that
$\gamma^\prime_0$ is equal to either $0$ or $1$ (our choice). From~(\ref{g0r}) we obtain
\begin{equation}
 P^2 = \left(\gamma_0 - \gamma^\prime_0\right) \frac{E_g (E_g + \Delta)}{E_g + \frac{2}{3}\Delta} \frac{\hbar^2}{2m_0},
\end{equation}
which we then use to modify the Luttinger parameters using~(\ref{g1r}-\ref{kr}). This method pushes unphysical solutions at large $\boldsymbol{k}$ out of interesting energies, whilst preserving the band structure around $\boldsymbol{k}=0$.

\begin{figure*}
\includegraphics[width=\textwidth]{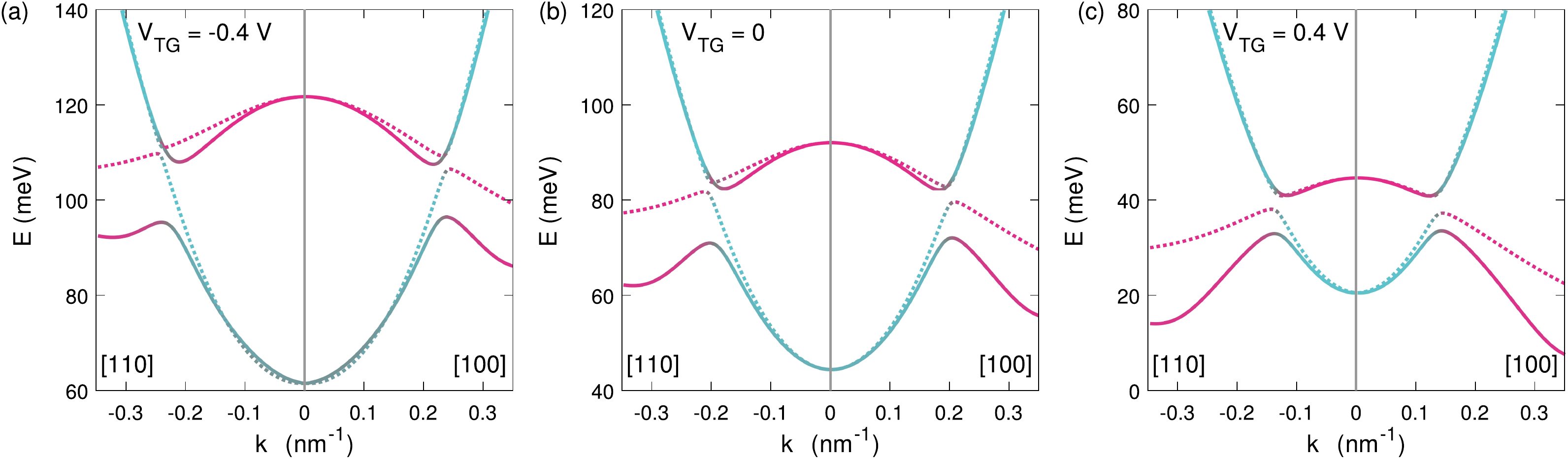}
\caption{Numerical band structure calculation of InAs/GaSb for two crystallographic directions at different top gate voltages. The band coloring represents the wavefunction character (blue for electron-like sates, red for hole-like states) while solid and dotted lines distinguish the two spin-orbit split bands. (a) $V_{TG}=-0.4~\mathrm{V}$, strong band splitting for the two lower bands and significant splitting in the two upper bands. (b) $V_{TG}=0$ (only built in electric field). Still strong spin-orbit splitting in the lower bands, still visible split in the upper bands. (c) $V_{TG}=0.4~\mathrm{V}$, significant split in the two lower bands, vanishing split in the upper bands.}
\label{disps}
\end{figure*}

For our simulations, we take the band structure parameters from~\cite{Halvorsen2000, Lawaetz1971} (summarized in Table~\ref{table_pars}). The valence band offsets~\cite{Halvorsen2000} are $0.56$ eV for GaSb-InAs, $0.18$ eV for AlSb-InAs and $-0.38$ eV for AlSb-GaSb.

\begin{table*}
\caption{\label{table_pars} Band structure parameters for InAs, GaSb and AlSb at $T=0$ K. (Ref.~\cite{Halvorsen2000, Lawaetz1971})}
\begin{ruledtabular}
\begin{tabular}{c c c c c c c c c}
& $E_g$ [eV] & $\Delta$ [eV]& $E_P$ [eV]& $m_c/m_0$ & $\gamma_1$ & $\gamma_2$ & $\gamma_3$ & $\kappa$ \\
InAs & 0.41 & 0.38 & 22.2 & 0.024 & 19.67 & 8.37 & 9.29 & 7.68 \\
GaSb & 0.8128 & 0.752 & 22.4 & 0.042 & 11.8 & 4.03 & 5.26 & 3.18 \\
AlSb & 2.32 & 0.75 & 18.7 & 0.18 & 4.15 & 1.01 & 1.75 & 0.31 \\
\end{tabular}
\end{ruledtabular}
\end{table*}

\subsubsection{Gate dependence of the band structure}

Computing the gate dependence of the band structure of InAs/GaSb quantum wells quantitatively requires a self-consistent solution of the $8\times 8$ Kane Hamiltonian and the Poisson equation. This problem involves both electron and hole densities, and while some approximate approaches have been discussed \cite{Andlauer2009}, it has not yet been solved satisfactorily.
For this reason we choose to only take into account the electrostatics due to gating on the level of the capacitor model. We extract a electrostatic potential $V(z)$, approximating the voltage drop between the nodes of the capacitor model as linear (this is justified as the dielectric constants of the different materials differ by at most a factor of 2). This potential enters the Kane Hamiltonian \eqref{H8x8} as an additional term on the diagonal. Finally, the spin texture shown in Fig.~1(c) of the main text are calculated as the expectation value of the electron spin Pauli matrices $\sigma_{x,y}$ at a constant energy.

The use of the capacitor model is justified in that we only strive to capture the qualitative aspects of the band structure. Further, as we see below, the spin-orbit features depend little on gating, as they are dominated by the intrinsic inversion symmetry breaking in the quantum well structure.

We present result of the band structure calculation for different top gate voltage in Fig.~\ref{disps}. We show results for the [110] crystallographic direction on the left hand side of each plot, and results for the [100] direction of the right hand side. The colors of the band indicate the wavefunction character (blue for electron-like and red for hole-like states) while solid and dashed lines distinguish the two spin-orbit split bands. We describe the proceude used to calculate the colors in Sec.~\ref{mixing}.
The biggest effect of the gate voltage is an change of the $k=0$ gap between the hole and the electron bands due to the electric field. As a consequence, the hybridization gap occurs at larger momenta for more negative top-gate voltages. At the same time, the hybridization gap becomes smaller and eventually vanishes. This is consistent with the experiment that finds still a significant residual conductance at the charge neutrality point.

The spin splitting in the band above the hybridization gap is gate-voltage dependent (from well visible at $V_{TG}=-0.4~\mathrm{V})$ to nearly vanishing at $V_{TG}=0.4~\mathrm{V})$). On the other hand, the large spin-orbit splitting in the bands below the hybridization gap is largely independent of gate voltage. Therefore at every gate voltage we can choose a Fermi level that corresponds to a system characterized by a single electron spin species. The hybridization gap also changes with gate voltage. For example a large positive gap is obtained for $V_{TG}=0.4~\mathrm{V})$. In that situation, and with the Fermi energy placed in the gap with the help of a back gate voltage, the system would reach the topological insulator regime.
As the top gate voltage is made more negative, the bands overlap increases and the hybridization gap reduces. In particular, already at $V_{TG}=-0.4~\mathrm{V}$ close inspection reveals the gap is anisotropic and vanishes along the [110] direction. As a result, at the energy level II of Fig.~1(b) of the main text, four Fermi pockets centered along the [110] direction could be present. In the present work we believe disorder potential could largely smear these features, if actually present in our samples. Furthermore their large effective mass would make negligible their contribution in transport. For this reason their presence is ignored in the left hand side of Fig.~1(c) of the main text.

\subsection{Estimation of electron-hole mixing in momentum states}
\label{mixing}
The wave functions of momentum states that we present in Fig.~1(b) of the main text and in Fig.~\ref{disps} contain both electron and hole components. Assuming the order of different wave function components is in agreement with the Hamiltonian of Eq.~\ref{H8x8}, we define
\begin{eqnarray}
|\psi_e|^2 &=& \sum_{n=1}^{n=2} \int |\psi_n(x)|^2\,dx; \\
|\psi_h|^2 &=& \sum_{n=3}^{n=8} \int |\psi_n(x)|^2\,dx,
\end{eqnarray}
where
\begin{equation}
|\psi_e|^2 + |\psi_h|^2 = 1.
\end{equation}

The blue color on the band structure plots corresponds to pure electron state ,$|\psi_e|^2=1$, and the pink color corresponds to pure hole state, $|\psi_h|^2=1$. The smooth color variation from blue to red indicates the mixing of electron and holes states along the energy bands.

\end{document}